\documentclass{article}

\usepackage{microtype}
\usepackage{graphicx}
\usepackage{subcaption}
\usepackage{booktabs}

\usepackage{hyperref}

\usepackage[preprint]{icml2026}

\usepackage{amsmath}
\usepackage{amssymb}
\usepackage{mathtools}
\usepackage{amsthm}

\usepackage[capitalize,noabbrev]{cleveref}

\usepackage[table]{xcolor}
\usepackage{colortbl}
\usepackage{tcolorbox}
\tcbuselibrary{breakable,skins}
\usepackage{tabularx}
\usepackage{makecell}
\usepackage{array}
\usepackage{enumitem}
\usepackage[textsize=tiny]{todonotes}

\newcolumntype{C}{>{\centering\arraybackslash}X}

\definecolor{headercolor}{RGB}{255, 242, 215}
\definecolor{rowcolor}{RGB}{245, 245, 245}
\definecolor{commentblue}{RGB}{0, 0, 255}
\definecolor{headerorange}{RGB}{255, 235, 210}
\definecolor{roworange}{RGB}{255, 248, 240}
\definecolor{headeryellow}{RGB}{255, 248, 220}
\definecolor{domainyellow}{RGB}{255, 252, 240}
\definecolor{rowgrey}{RGB}{248, 248, 248}

\newtcolorbox{promptbox}{
  enhanced,
  breakable,
  colback=gray!3,
  colframe=magenta!80,
  boxrule=0.6pt,
  arc=2pt,
  left=6pt,
  right=6pt,
  top=6pt,
  bottom=6pt,
  fonttitle=\bfseries,
  title=Meta-Agent System Prompt,
  center title,
  label={box:meta-agent-prompt}
}

\theoremstyle{plain}

\theoremstyle{definition}

\theoremstyle{remark}

\icmltitlerunning{PhysicsAgentABM: Physics-Guided Generative Agent-Based Modeling}

\begin{document}

\twocolumn[
  \icmltitle{PhysicsAgentABM: Physics-Guided Generative Agent-Based Modeling}


  \begin{icmlauthorlist}
    \icmlauthor{Kavana Venkatesh}{vt}
    \icmlauthor{Yinhan He}{uva}
    \icmlauthor{Jundong Li}{uva}
    \icmlauthor{Jiaming Cui}{vt}
  \end{icmlauthorlist}

  \icmlaffiliation{vt}{Virginia Tech}
  \icmlaffiliation{uva}{University of Virginia}

  \icmlcorrespondingauthor{Kavana Venkatesh}{kavanav@vt.edu}
  \icmlcorrespondingauthor{Jiaming Cui}{jiamingcui@vt.edu}

  \vskip 0.3in
]

\printAffiliationsAndNotice{}  

\begin{abstract}
Large language model (LLM)-based multi-agent systems enable expressive agent reasoning but are expensive to scale and poorly calibrated for timestep-aligned state-transition simulation, while classical agent-based models (ABMs) offer interpretability but struggle to integrate rich individual-level signals and non-stationary behaviors. We propose \texttt{PhysicsAgentABM}, which shifts inference to behaviorally coherent agent clusters: state-specialized symbolic agents encode mechanistic transition priors, a multimodal neural transition model captures temporal and interaction dynamics, and uncertainty-aware epistemic fusion yields calibrated cluster-level transition distributions. Individual agents then stochastically realize transitions under local constraints, \textit{decoupling} population inference from entity-level variability. We further introduce \texttt{ANCHOR}, an LLM agent-driven clustering strategy based on cross-contextual behavioral responses and a novel contrastive loss, reducing LLM calls by up to $\sim$6--8$\times$. Experiments across public health, finance, and social sciences show consistent gains in event-time accuracy and calibration over mechanistic, neural, and LLM baselines. By re-architecting generative ABM around population-level inference with uncertainty-aware neuro-symbolic fusion, \texttt{PhysicsAgentABM} establishes a new paradigm for scalable and calibrated simulation with LLMs.
\end{abstract}

\section{Introduction}
\label{sec:intro}

\begin{figure*}[t]
    \centering
    \includegraphics[width=\textwidth]{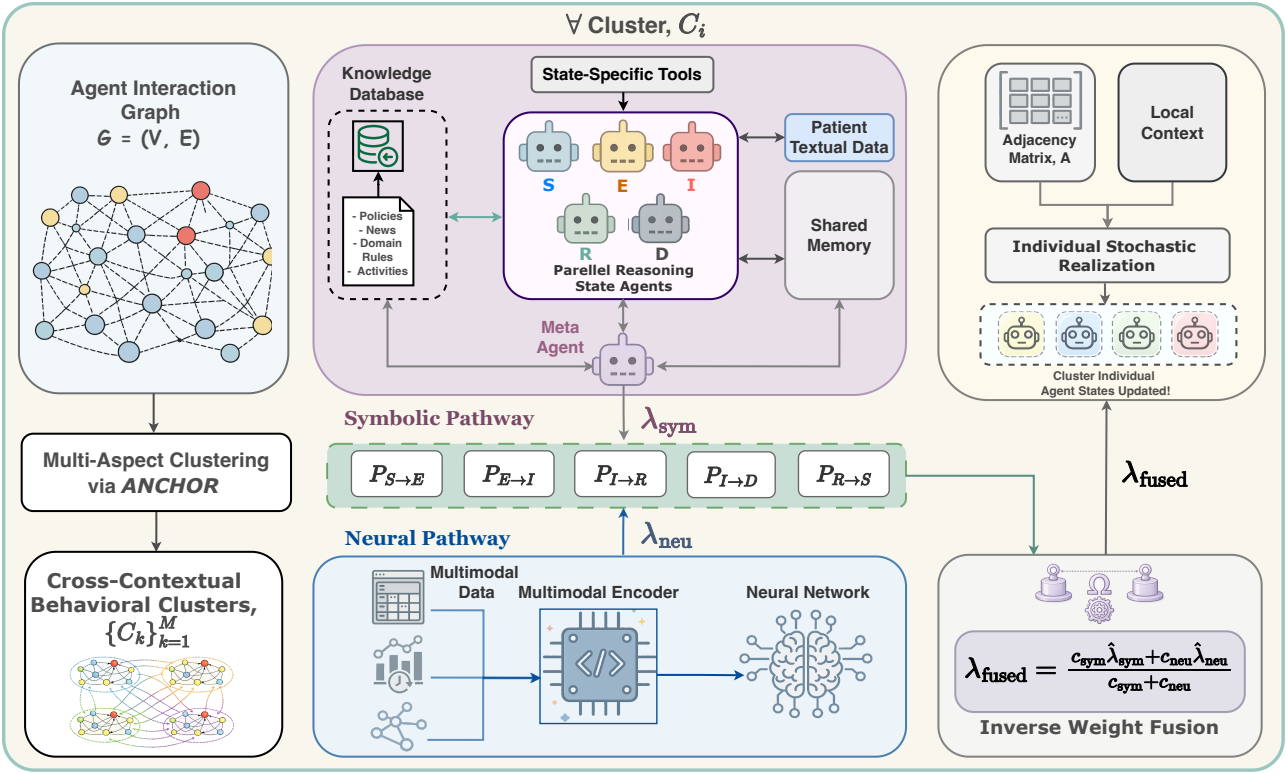}
    \caption{
\textbf{Overview of PhysicsAgentABM Architecture.}
Inference is performed at the cluster level via symbolic and neural pathways with uncertainty-aware fusion, followed by stochastic agent-level realization. ANCHOR enables behaviorally coherent abstraction.
}
\label{fig:architecture-overview}
\end{figure*}

Many complex systems of scientific and societal importance, such as infectious disease spread \cite{deng2020cola,pastor2015epidemic}, social diffusion \cite{qiu2018deepinf,kempe2003maximizing}, financial contagion \cite{xu2021hist,xiang2022temporal}, and infrastructure failure cascades \cite{kipf2018neural,fang2021spatial} are governed by state-dependent interactions unfolding across multiple scales \cite{velivckovic2017graph,battaglia2018relational}. Individual entities do not evolve in isolation: their behaviors are shaped by local interactions, shared group-level influences, and broader contextual signals such as policies, norms, or collective risk perception \cite{ying2018hierarchical,lowe2017multi}. Accurately simulating such systems therefore requires models that can jointly capture mechanistic structure, data-driven dynamics, symbolic context, and uncertainty while remaining scalable and interpretable. 

Agent-based models (ABMs) provide a principled foundation by explicitly modeling interacting entities through mechanistic transition rules, yielding interpretability and emergent behavior \cite{epstein2012generative, bonabeau2002agent}. However, classical ABMs rely on hand-crafted, static rules and coarse calibration, limiting adaptation to heterogeneous populations, multimodal signals, and non-stationary regimes \cite{pastor2015epidemic}. Learned transition models, including probabilistic graphical models, neural state-space models, and graph neural networks have been incorporated to improve flexibility \cite{murphy2012machine, krishnan2017structured, kipf2017semi, hamilton2017inductive, rossi2020temporal}, but often obscure mechanistic structure and exhibit poor uncertainty calibration under distribution shift \cite{ovadia2019can}. More recently, large language models (LLMs) have enabled generative agent-based models (GABMs) with rich symbolic reasoning and memory \cite{park2023generative, generative1000people2024, socioverse2025}. Parallel lines of work integrate LLMs into ABM pipelines for social simulation and conversational dynamics \cite{llmempowerabm2023, abmgenai2024, agentsociety2025}. Yet most LLM-based GABMs perform reasoning at the level of individual agents, incurring high computational cost, weak grounding in relational data, and unreliable stochastic behavior in the absence of principled uncertainty modeling \cite{llmtrust2024, limitsagency2024}.

Taken together, existing approaches reveal two fundamental gaps in generative agent-based modeling. First, inference is performed at the level of individual agents \cite{park2023generative,gao2024agentscope} despite strong group-level, institutional, and contextual forces that drive coherent population dynamics \cite{granovetter1978threshold,centola2018behavior}. Second, symbolic reasoning and neural learning are combined heuristically \cite{manhaeve2018deepproblog,yi2018neural} rather than treated as complementary epistemic sources with explicit uncertainty modeling. Real-world systems make these gaps explicit: epidemic dynamics shift coherently within communities \cite{chang2021mobility,block2020social}, financial institutions co-move through shared exposures \cite{billio2012econometric,diebold2014network}, and social groups align around common narratives. Capturing such structure requires population-level inference that produces calibrated priors for individual realization \cite{gelman1995bayesian,ovadia2019can}, rather than isolated agent simulation.

We introduce \texttt{PhysicsAgentABM}, a hierarchical neuro-symbolic framework that addresses these challenges by redefining inference in generative agent-based models. \texttt{PhysicsAgentABM} elevates inference from individual agents to adaptive agent clusters, where shared dynamics, contextual influences, and uncertainty are modeled explicitly. Cluster-level predictions define probabilistic transition priors \cite{tran2017hierarchical,fortuin2019deep}, while individual agents stochastically realize state transitions conditioned on local attributes and neighborhood context, preserving heterogeneity without sacrificing population-level coherence (Figure~\ref{fig:architecture-overview}). Within each cluster, a state-specialized symbolic reasoning layer coordinated by a meta-agent encodes mechanistic constraints and regime context, while a multimodal neural transition model \cite{baltruvsaitis2018multimodal} captures temporal and interaction-driven regularities. These pathways are treated as distinct epistemic hypotheses and reconciled through uncertainty-aware fusion \cite{hullermeier2021aleatoric} to yield calibrated, population-consistent dynamics.
\begin{figure*}[t]
    \centering
    \includegraphics[width=\textwidth]{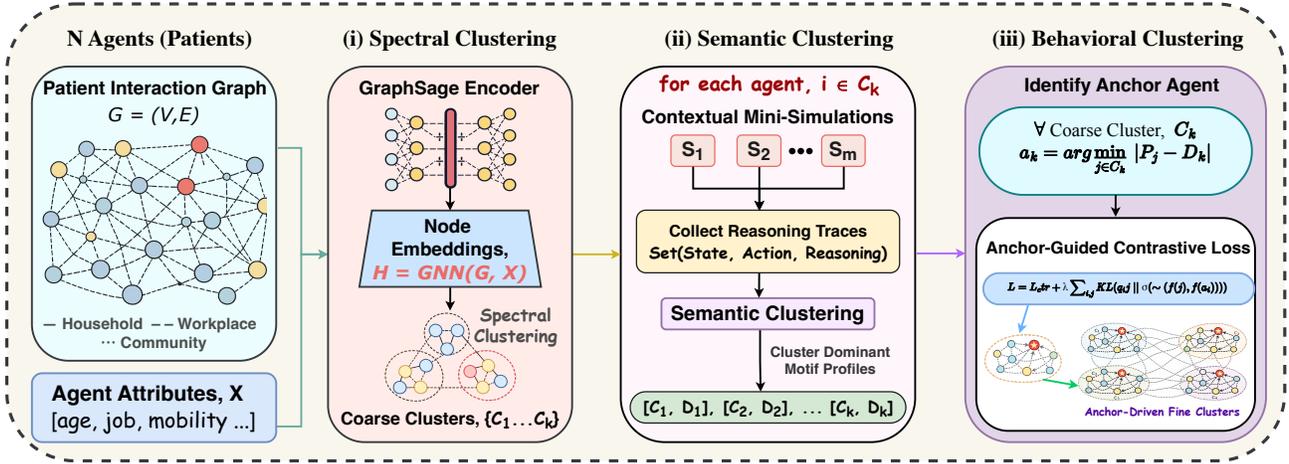}
    \caption{
\textbf{ANCHOR Overview.}
An overview of our clustering mechanism. 
}
    \label{fig:anchor-overview}
\end{figure*}
A critical enabler of this hierarchy is clustering itself. In generative agent-based models, clustering must be semantically meaningful, transition-faithful, and adaptive to evolving dynamics. Classical graph clustering methods optimize structural criteria while ignoring behavioral and dynamical semantics \cite{schaeffer2007graph,loukas2019graph}, whereas existing LLM-based simulators \cite{park2023generative,hong2023metagpt,qian2024chatdev} use LLMs to execute agent behavior rather than to reason about abstraction. We therefore introduce \texttt{ANCHOR}, a novel LLM-agent-driven clustering mechanism in which symbolic agents act as semantic controllers of abstraction \cite{huang2022language}, organizing populations based on evolving cross-contextual behavioral similarity \cite{newman2006modularity,yang2009combining}, interaction structure, and state-transition tendencies under graph constraints. By using LLMs to control where and how abstraction occurs rather than to simulate individual agents, \texttt{ANCHOR} enables scalable symbolic reasoning aligned with downstream probabilistic transition modeling. Our contributions are threefold:
\vspace{-10pt}
\begin{itemize}
\item We introduce \texttt{PhysicsAgentABM}, a hierarchical neuro-symbolic framework for cluster-level inference via state-specialized symbolic and neural reasoning.
\vspace{-6pt}
\item We propose \texttt{ANCHOR}, the \textit{first} LLM-agent-anchored clustering method that treats abstraction as a semantic control problem for transition-faithful simulation.
\vspace{-6pt}
\item We introduce a calibrated, timestep-aligned paradigm that decouples population-level inference from agent-level realization for robust long-horizon dynamics.
\end{itemize}

\section{Related Work}
\label{sec:related}


\subsection{LLM-Based Multi-Agent Simulation Frameworks}
LLMs have enabled generative multi-agent systems with language-based reasoning and coordination \cite{wang2024survey, guo2024large}. Frameworks such as MetaGPT and ChatDev demonstrate role-specialized collaboration \cite{hong2023metagpt, qian2024chatdev}, while Generative Agents and AgentScope model large-scale social behavior \cite{park2023generative, gao2024agentscope}. Related work studies coordination protocols and grounding in structured environments \cite{li2023camel, wu2024autogen, zhu2023ghost, wang2023plan}. However, most approaches operate at the individual-agent level, incurring high computational cost and limiting scalability \cite{mou2024unveiling}. Their reliance on LLM priors without explicit mechanistic structure further limits interpretability, uncertainty quantification, and physical alignment, leading to temporal inconsistency and miscalibration under distribution shift \cite{du2023improving}.

\subsection{ABM for Social and Health Dynamics}
ABMs simulate complex social and health systems by modeling interacting individuals and emergent collective behavior \cite{bonabeau2002agent, epstein2012generative}. They have been widely applied to infectious disease dynamics \cite{hethcote2000mathematics, pastor2015epidemic}, social contagion and behavioral cascades \cite{granovetter1978threshold, watts2002simple}, and systemic risk propagation in networks \cite{hurd2016contagion, gai2010contagion}. Recent work incorporates large language models to enrich agent decision-making and simulate more realistic behavior \cite{park2023generative}. However, these approaches rely on hand-crafted rules or costly per-agent LLM inference, limiting principled integration of multimodal signals into population-level dynamics.

\subsection{Graph and Behavior-Aware Clustering Methods}
Graph clustering optimizes structural objectives such as modularity \cite{newman2006modularity}, spectral criteria \cite{von2007tutorial}, or edge density \cite{blondel2008fast}, but typically ignores node attributes, temporal dynamics, and behavioral semantics. Feature-aware extensions with differentiable pooling \cite{ying2018hierarchical} and temporal graph models \cite{rossi2020temporal} capture evolving structure, yet remain agnostic to decision-making behavior. Recent work applies large language models to semantic graph reasoning \cite{pan2024unifying, ye2024language}, but largely treats LLMs as embedding generators. In contrast, \texttt{ANCHOR} uses LLMs as \emph{semantic controllers of abstraction}, enabling transition-faithful clustering via cross-contextual behavioral probing aligned with downstream state-transition modeling.

\section{Methodology}
\label{sec:method}

\subsection{PhysicsAgentABM Architecture}
\label{sec:architecture}
We model the population as an interaction graph $\mathcal{G}=(\mathcal{V},\mathcal{E})$ with adjacency matrix $\mathbf{A}\in\{0,1\}^{N\times N}$, where each entity $i\in\mathcal{V}$ occupies a discrete state $x_i(t)\in\mathcal{S}=\{S_1,\dots,S_K\}$. The state space $\mathcal{S}$ is application-dependent (e.g., \{\emph{susceptible, exposed, infected, recovered, deceased}\} in epidemiology or \{\emph{bullish, bearish, neutral}\} in finance), and $\mathbf{A}$ encodes local mechanistic interactions such as contact or influence. For scalability, inference relies on neighborhood statistics derived from $\mathbf{A}$ rather than explicit pairwise modeling. Entities are partitioned into $M$ clusters $\{C_k\}_{k=1}^M$ of agents exhibiting similar behavioral responses to contextual stimuli, constructed via ANCHOR (Section~\ref{sec:anchor}). Within each cluster $C_k$, \texttt{PhysicsAgentABM} infers population-level state transition dynamics via two complementary pathways: symbolic reasoning and neural prediction, both estimating hazards over the same set of valid transitions $\mathcal{T}$ but with distinct inductive biases. The symbolic pathway operates on a cluster-level context
$\mathcal{C}_k(t)=\big(\phi_k(t),\psi_k(t),\psi_k^{\pm}(t)\big)$,
where $\phi_k(t)$ summarizes state composition, $\psi_k(t)$ captures exogenous temporal context, and $\psi_k^{\pm}(t)$ aggregates neighboring-cluster signals. Conditioned on $\mathcal{C}_k(t)$, a meta-agent coordinates state-specialized agents to estimate symbolic hazards
$\Lambda^{\text{sym}}_k(t)=\{\hat{\lambda}^{k,\text{sym}}_{s\rightarrow s'}(t)\}_{(s,s')\in\mathcal{T}}$
with associated epistemic uncertainty $u^k_{s\rightarrow s'}(t)$, modeling population-level transition tendencies rather than discrete actions. In parallel, the neural pathway estimates cluster-level hazards from aggregated multimodal inputs
$\mathbf{x}_k(t)=[\text{tabular}_k(t),\ \text{temporal}_k(t),\ \text{graph}_k]$
using a multimodal encoder to produce $\hat{\Lambda}^{\text{neu}}_k(t)$. Symbolic and neural estimates are combined via uncertainty-aware epistemic fusion at the transition level:
\begin{equation}
\lambda^{\text{fused}}
=
\frac{
c^{\text{sym}}\,\hat{\lambda}^{\text{sym}}
+
c^{\text{neu}}\,\hat{\lambda}^{\text{neu}}
}{
c^{\text{sym}} + c^{\text{neu}}
},
\end{equation}
where confidence terms $(c^{\text{sym}},c^{\text{neu}})$ are adaptively calibrated by a lightweight MLP. The resulting fused hazards define population-consistent transition priors. Individual entities then stochastically realize state transitions by modulating these priors with local attributes and neighborhood statistics from $\mathbf{A}$, decoupling population-level inference from agent-level realization (Section~\ref{sec:decoupled-simulation}). See Figure \ref{fig:architecture-overview} and Algorithm \ref{alg:PhysicsAgentABM} for an overview of our framework.

\begin{algorithm}[t]
\caption{\textbf{PhysicsAgentABM Simulation Workflow}}
\label{alg:PhysicsAgentABM}
\small
\textbf{Input:} Interaction graph $\mathcal{G}=(\mathcal{V},\mathcal{E})$, adjacency $\mathbf{A}$,
clusters $\{C_k\}_{k=1}^M$,
state space $\mathcal{S}$,
transition set $\mathcal{T}$,
time horizon $T$ \\
\textbf{Output:} Agent state trajectories $\{x_i(t)\}_{i=1,t=1}^{N,T}$

\begin{algorithmic}[1]

\FOR{each agent $i \in \mathcal{V}$}
    \STATE Initialize state $x_i(0) \in \mathcal{S}$ and memory $\mathcal{M}_i$
\ENDFOR

\FOR{$t = 1$ to $T$}

    \FOR{each cluster $C_k$}
        \STATE Compute regime context $\mathcal{C}_k(t) = (\phi_k(t), \psi_k(t), \psi_k^{\pm}(t))$
        \FOR{each transition $(s \rightarrow s') \in \mathcal{T}$}
            \STATE $\hat{\lambda}^{k,\text{sym}}_{s\rightarrow s'}(t) \leftarrow \mathcal{A}^{\text{sym}}_{k,s}(\mathcal{C}_k(t))$
            \STATE $\hat{\lambda}^{k,\text{neu}}_{s\rightarrow s'}(t) \leftarrow f_{\theta}(\mathbf{x}_k(t))$
            \STATE $\lambda^{k}_{s\rightarrow s'}(t) \leftarrow 
            \mathrm{Fuse}\!\left(\hat{\lambda}^{k,\text{sym}}_{s\rightarrow s'}(t),
            \hat{\lambda}^{k,\text{neu}}_{s\rightarrow s'}(t)\right)$
        \ENDFOR
    \ENDFOR

    \FOR{each agent $i \in \mathcal{V}$}
        \STATE Identify cluster $C_k$ s.t. $i \in C_k$
        \FOR{each outgoing transition $(s \rightarrow s')$ from $x_i(t)=s$}
            \STATE $\tilde{\lambda}^{i}_{s\rightarrow s'}(t) \leftarrow 
            g\!\left(\lambda^{k}_{s\rightarrow s'}(t), \mathcal{M}_i, \mathbf{A}_i \right)$
        \ENDFOR
        \STATE $x_i(t+1) \sim \mathrm{Categorical}(\{\tilde{\lambda}^i_{s\rightarrow s'}(t)\}_{s'\in\mathcal{S}_i}\cup\{1\})$
    \ENDFOR

\ENDFOR

\STATE \textbf{return} $\{x_i(t)\}_{i,t}$

\end{algorithmic}
\end{algorithm}


\subsection{ANCHOR: Agent-Driven Multi-Stage Clustering}
\label{sec:anchor}

\textbf{Stage 1: Structural-Semantic Initialization.}
Given an interaction graph $G=(V,E)$ with edge weights $W_{ij}$ and agent attributes $X$, \texttt{ANCHOR} computes scalable structural embeddings $H \leftarrow \textsc{GraphSAGE}(G,X)$ capturing local topology and neighborhood semantics. Embeddings are then concatenated with attributes to form $Y=[H\parallel X]$. Spectral clustering  on $Y$ yields coarse clusters $\{C_i\}_{i=1}^{K_{\text{coarse}}}$, providing a structural prior without invoking agent reasoning.

\textbf{Stage 2: Behavioral Motif Discovery and Agent Profiling.}

To capture decision-making behavior beyond structure, we run short-horizon domain-specific mini-simulations under controlled diagnostic scenarios $S=\{s_1,\ldots,s_M\}$. For each agent $j$ and scenario $s$, we record reasoning-action traces $(s,r_j^s,a_j^s)$, which are embedded and clustered to identify behavioral motifs $\{M_k\}_{k=1}^{K_m}$. A behavioral motif is a recurring pattern in how an agent responds to seemingly diverse situations in a given problem setting (mobility at home vs community). Each agent is summarized by a motif frequency profile
$P_j=[\mathrm{freq}(M_1),\ldots,\mathrm{freq}(M_{K_m})]$, 
and each coarse cluster by its dominant motif profile
$D_i=\frac{1}{|C_i|}\sum_{j\in C_i}P_j$.
This representation decouples behavioral control from topology and forms the basis for `anchor agent' selection and contrastive refinement in the next stage.

\begin{figure}[t]
    \centering
    \includegraphics[width=\columnwidth]{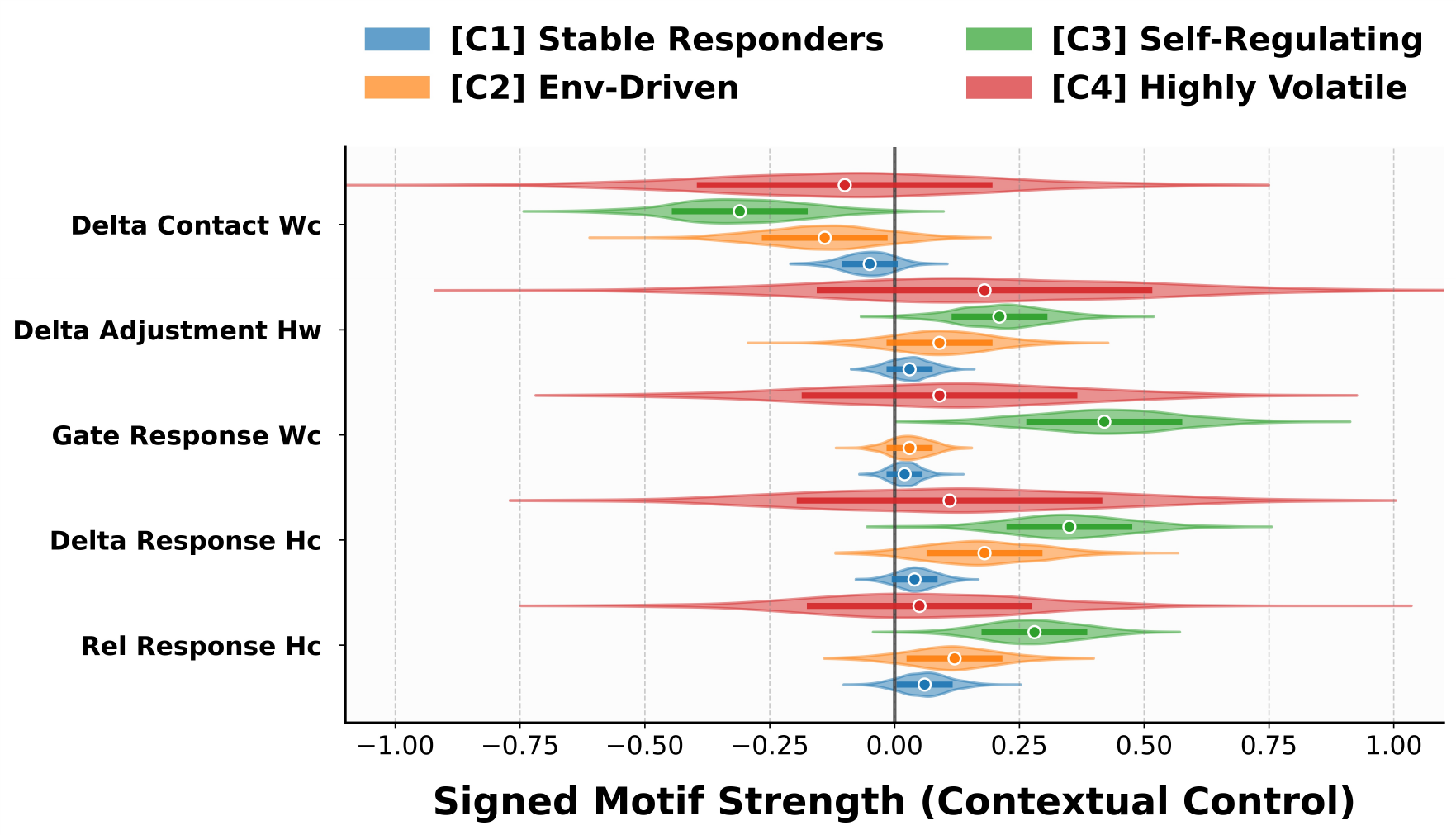}
    \caption{
\textbf{ANCHOR cross-contextual clusters (epidemiology).}
\texttt{ANCHOR} identifies four distinct clusters among 1,000 agents based on joint semantic and cross-contextual behavioral responses. We report normalized averages of the top-5 dominant behavioral motifs per cluster, capturing coordinated shifts in response intensity, isolation, and compliance across contexts, with signed values indicating motif direction and strength. See Appendix Section~\ref{supp-sec:anchor} for detailed motif descriptions and interpretation.}
\label{fig:anchor-intepretation}
\end{figure}

\textbf{Stage 3: Anchor-Guided Contrastive Refinement.}
For each coarse cluster $C_i$, we select an anchor agent
$a_i=\arg\min_{j\in C_i}\|P_j-D_i\|_2$
as a behaviorally representative reference in the motif space from Stage~2. Anchors actively guide refinement by evaluating regime compatibility: for each anchor–agent pair $(a_i,j)$, the anchor produces a soft judgment $q_{ij}\in[0,1]$ via LLM-based reasoning over behavioral motif summaries and contextual probes, indicating whether agent $j$ follows decision logic consistent with the anchor’s regime. Each agent $j$ is encoded as $f(j)=[H_j\parallel P_j\parallel \text{context}_j]$. Clustering is refined by optimizing a novel joint objective that couples contrastive learning with anchor judgments:
\begin{equation}
\mathcal{L}
=
\mathcal{L}_{\text{ctr}}
+
\lambda
\sum_{i,j}
\mathrm{KL}
\Big(
q_{ij}
\;\Big\|\;
\sigma\big(\mathrm{sim}(f(j),f(a_i))\big)
\Big),
\end{equation}
where $\mathcal{L}_{\text{ctr}}$ enforces regime separation and the alignment term matches learned similarity to anchor-defined compatibility. We set $\lambda=1/|\mathcal{N}|$, where $N$ is the number of agent–anchor comparisons in the batch. Final representations $Z_j=\alpha H_j+\beta f(j)+\gamma P_j$ with $\alpha+\beta+\gamma=1$ are learned to maximize cluster quality, and hierarchical clustering on $\{Z_j\}$ yields refined clusters aligned by functional decision logic and behavioral similarity, even across topologically distant agents. See Fig. \ref{fig:anchor-overview} for an overview.

\textbf{Stage 4: Boundary Optimization and Adaptation:}
Boundary agents near multiple cluster centroids are reassigned using a motif-guided pull score 
$\mathrm{pull}(j, C_i) = \cos(P_j, D_i) \cdot \mathrm{conn}(j, C_i)$, where $\mathrm{conn}(j, C_i)$ denotes the normalized fraction of agent $j$’s neighbors belonging to cluster $C_i$. ANCHOR merges behaviorally redundant clusters and splits heterogeneous ones based on motif divergence and entropy. Cluster quality is monitored via modularity, silhouette score, and motif coherence, which tracks the consistency of local interaction motifs within clusters across refinement steps; refinement is re-invoked upon degradation. See additional results and full algorithm in Section \ref{supp-sec:anchor}.

\subsection{Decoupled Population-to-Entity Simulation}
\label{sec:decoupled-simulation}

Using cluster-level transition hazards from Sec.~\ref{sec:architecture}, Entity Agents stochastically realize individual state transitions under local constraints. For an agent $i$ in cluster $C_k$ occupying state $s$ at time $t$, the fused hazard $\lambda^k_{s\rightarrow s'}(t)$ encodes the population-consistent transition tendency and serves as a base rate. This hazard is modulated by agent-specific attributes and neighborhood interactions from the adjacency matrix $\mathbf{A}$, yielding an individualized transition intensity $\tilde{\lambda}^i_{s\rightarrow s'}(t)$ without additional LLM calls (at most one inference per cluster). Each agent samples its next state via a competing-risk realization, $x_i(t+1) \sim \mathrm{Categorical}(\{\tilde{\lambda}^i_{s\rightarrow s'}(t)\}_{s'\in\mathcal{S}_i}\cup\{1\})$. The distribution is normalized, with the final mass corresponding to temporal persistence. Competing transitions (e.g., recovery vs.\ death) are treated as concurrent risks whose relative intensities determine the realized event, while temporal persistence corresponds to survival in the absence of any transition. This synchronous realization preserves individual heterogeneity while enforcing a shared population regime, consistent with standard stochastic epidemic and agent-based modeling practice \cite{gillespie1977exact,pastor2015epidemic}.

\section{Experiments}
\label{sec:experiments}
\subsection{Experimental Setup}
\label{sec:exp-setup}

We evaluate \texttt{PhysicsAgentABM} on three partially observable domains with regime shifts and latent population dynamics: \textbf{epidemiology}, \textbf{finance}, and \textbf{social attention}. In all settings, agents interact over structured networks, evolve through discrete latent states, and are evaluated against population-level ground truth signals. \textbf{Epidemiology:} Singapore MOH COVID-19 data \cite{owid-coronavirus} (first 1,000 confirmed cases; Jan 23-Apr 14, 2020) with infection/recovery/death timestamps and contact tracing; ground-truth SEIRD states are inferred using a 7-day pre-symptomatic exposure window. Agents interact via a multi-layer contact network (household/workplace/community), and cluster-level neural inputs aggregate demographics, latent state proportions, transition statistics, network embeddings, and exogenous policy/mobility signals.
\textbf{Finance:} Market sentiment diffusion with a synthetic population of 100 traders (heterogeneous wealth, risk tolerance, trading styles, portfolios) over the top 20 S\&P 500 stocks across two financial quarters (Jul 1-Dec 31, 2024), interacting through a correlation network from historical returns. Since individual sentiment states (\emph{Bullish/Bearish/Neutral}) are unobserved, population-level ground truth uses daily S\&P 500 market regimes as a proxy for collective risk appetite. \textbf{Attention Lifecycle:} Population attention to \emph{Climate Change} over 90 days (Dec 1, 2024–Feb 28, 2025) with 250 agents (heterogeneous activity and topic affinity) interacting over a social graph; latent states (\emph{Unaware/Interested/Fatigued}) are evaluated using normalized daily Wiki page views to capture emergence, amplification, and fatigue. Full details are in Appendix Sec.~\ref{supp-sec:dataset-construction}, \ref{supp-sec:neural-pathway} and \ref{supp-sec:symblic-reasoning}.

\begin{table*}[t]
\centering
\caption{
\textbf{Agent-level state transition evaluation across domains.}
Rolling-window evaluation with 7-day test windows. For each metric, in epidemiology, agent SEIRD predictions are evaluated against individual-level states inferred from Singapore MOH data; in finance, agent sentiment states are compared to daily S\&P 500 market regimes as a population-level signal; in social attention, predicted agent states are aggregated and evaluated against normalized daily Wiki pageviews. \textit{DeepProbLog} excludes EETE as it predicts event types without explicit event-time distributions.}
\label{tab:eval-quant-general}
\footnotesize
\setlength{\tabcolsep}{3.5pt}
\renewcommand{\arraystretch}{1.3} 

\rowcolors{3}{}{rowcolor}

\begin{tabular}{l|cc|cc|cc|cc|c}
\toprule
\rowcolor{headerorange} 
Metric
& \multicolumn{2}{c|}{\textit{Mechanistic}}
& \multicolumn{2}{c|}{\textit{Pure Neural}}
& \multicolumn{2}{c|}{\textit{LLM-based}}
& \multicolumn{2}{c|}{\textit{Hybrid}}
& \textbf{Ours} \\

\rowcolor{headerorange}
& Rule-ABM & MF-Markov
& GNN-LSTM & TGN
& LLM-Agent & LLM-MAS
& DeepProbLog & Rule-NN & \\

\specialrule{\lightrulewidth}{0pt}{0pt}
\rowcolor{roworange}
\multicolumn{10}{c}{\rule{0pt}{3ex} \textbf{Epidemiology} \rule[-1.2ex]{0pt}{0pt}} \\
\specialrule{\lightrulewidth}{0pt}{0pt}

EETE$\downarrow$
& 3.75$\pm$0.09 & 4.61$\pm$0.05  
& 4.35$\pm$0.04 & 3.25$\pm$0.03
& 3.95$\pm$0.06 & 3.37$\pm$0.07
& N/A & 5.12$\pm$0.03
& \textbf{1.92$\pm$0.05} \\

ET-F1$\uparrow$
& 0.45$\pm$0.04 & 0.33$\pm$0.01     
& 0.49$\pm$0.09 & 0.59$\pm$0.05
& 0.57$\pm$0.07 & 0.62$\pm$0.09
& 0.51$\pm$0.05 & 0.24$\pm$0.04
& \textbf{0.81$\pm$0.01} \\

NLL$\downarrow$
& 1.96$\pm$0.07 & 3.02$\pm$0.08  
& 2.81$\pm$0.05 & 1.72$\pm$0.04
& 1.88$\pm$0.09 & 1.27$\pm$0.10
& 2.11$\pm$0.05 & 19.04$\pm$0.08
& \textbf{0.73$\pm$0.03} \\

Brier$\downarrow$
& 0.86$\pm$0.06 & 0.95$\pm$0.03  
& 0.78$\pm$0.03 & 0.68$\pm$0.02
& 0.82$\pm$0.06 & 0.65$\pm$0.07
& 0.90$\pm$0.03 & 0.96$\pm$0.02
& \textbf{0.16$\pm$0.01} \\

\specialrule{\lightrulewidth}{0pt}{0pt}
\rowcolor{roworange}
\multicolumn{10}{c}{\rule{0pt}{3ex} \textbf{Financial Contagion} \rule[-1.2ex]{0pt}{0pt}} \\
\specialrule{\lightrulewidth}{0pt}{0pt}

EETE$\downarrow$
& 5.3$\pm$0.07 & 3.22$\pm$0.06
& 3.85$\pm$0.06 & 3.29$\pm$0.05
& 3.22$\pm$0.08 & 2.73$\pm$0.09
& N/A & 3.31$\pm$0.05
& \textbf{2.35$\pm$0.04} \\

ET-F1$\uparrow$
& 0.42$\pm$0.06 & 0.27$\pm$0.01
& 0.28$\pm$0.02 & 0.49$\pm$0.01
& 0.65$\pm$0.09 & 0.71$\pm$0.11
& 0.56$\pm$0.07 & 0.54$\pm$0.05
& \textbf{0.76$\pm$0.04} \\

NLL$\downarrow$
& 1.20$\pm$0.10 & 1.05$\pm$0.08
& 1.04$\pm$0.06 & 0.68$\pm$0.06
& 0.36$\pm$0.11 & 0.29$\pm$0.12
& 1.08$\pm$0.07 & 0.82$\pm$0.06
& \textbf{0.49$\pm$0.04} \\

Brier$\downarrow$
& 0.86$\pm$0.05 & 0.79$\pm$0.04
& 0.88$\pm$0.05 & 0.73$\pm$0.04
& 0.27$\pm$0.08 & 0.29$\pm$0.09
& 0.69$\pm$0.05 & 0.55$\pm$0.03
& \textbf{0.22$\pm$0.02} \\

\specialrule{\lightrulewidth}{0pt}{0pt}
\rowcolor{roworange}
\multicolumn{10}{c}{\rule{0pt}{3ex} \textbf{Social Diffusion} \rule[-1.2ex]{0pt}{0pt}} \\
\specialrule{\lightrulewidth}{0pt}{0pt}

EETE$\downarrow$
& 3.87$\pm$0.06 & 3.41$\pm$0.09
& 3.33$\pm$0.02 & 3.30$\pm$0.03
& 4.38$\pm$0.01 & 3.13$\pm$0.08
& N/A & 3.19$\pm$0.06
& \textbf{2.48$\pm$0.02} \\

ET-F1$\uparrow$
& 0.48$\pm$0.09 & 0.29$\pm$0.01
& 0.31$\pm$0.05 & 0.44$\pm$0.04
& 0.47$\pm$0.08 & 0.48$\pm$1.00
& 0.33$\pm$0.06 & 0.27$\pm$0.04
& \textbf{0.64$\pm$0.09} \\

NLL$\downarrow$
& 1.95$\pm$0.04 & 1.87$\pm$0.06
& 1.69$\pm$0.06 & 1.63$\pm$0.05
& 2.91$\pm$0.01 & 1.94$\pm$0.08
& 2.66$\pm$0.06 & 3.58$\pm$0.05
& \textbf{0.59$\pm$0.03} \\

Brier$\downarrow$
& 0.64$\pm$0.03 & 0.81$\pm$0.09
& 0.83$\pm$0.01 & 0.74$\pm$0.05
& 0.45$\pm$0.03 & 0.36$\pm$0.02
& 0.88$\pm$0.07 & 0.82$\pm$0.07
& \textbf{0.12$\pm$0.07} \\

\bottomrule
\end{tabular}
\end{table*}

\textbf{Baselines:} We compare \texttt{PhysicsAgentABM} against eight baselines spanning four paradigms.
\emph{Mechanistic:} Rule-ABM \cite{terHoeven2025Mesa3} and MF-Markov \cite{mckean1966class}.
\emph{Neural:} GNN-LSTM \cite{scarselli2008graph, hochreiter1997long} and TGN \cite{rossi2020temporal}.
\emph{LLM-based:} a single LLM-Agent and a flat LLM multi-agent system \cite{park2023generative}.
\emph{Hybrid:} DeepProbLog \cite{manhaeve2018deepproblog} and Rule-NN \cite{andrews1995survey}. Baselines operate at the individual-agent level without hierarchical abstraction or explicit uncertainty fusion. For fairness, all methods are evaluated under the same data availability and supervision regime, with inputs adapted to each model class (mechanistic, neural, or LLM-based).

\textbf{Evaluation Protocol:} We adopt a rolling-window forecasting protocol to reflect realistic deployment and prevent temporal leakage. Each window uses a 28-day lookback for training or calibration followed by a 7-day forecast horizon, with models re-trained or re-calibrated at every step. Metrics are computed only on event-bearing trajectories within the forecast window, avoiding trivial gains from persistent terminal states. Evaluation emphasizes event timing, regime coherence, and calibration, rather than pointwise accuracy, testing whether models recover population-level dynamics such as emergence, propagation, and attenuation under partial observability and non-stationarity (e.g., policy shifts, market regime changes, viral events). We report both quantitative metrics and qualitative trajectory analyses. Initial states follow empirical first infection day state distributions for epidemiology (e.g., January 23, 2020 at ${S{:}991,E{:}8,I{:}1}$), with a 1\% minority-state initialization in other domains to reflect realistic early-phase conditions where rare-state emergence drives downstream dynamics. See appendix Sec. \ref{supp-sec:dataset-construction}.

\textbf{Evaluation Metrics:} Standard pointwise metrics such as MAE or accuracy are insufficient for agent-based state-transition modeling, as they ignore event timing, collapse temporal uncertainty, and are dominated by persistent terminal states in long-horizon simulations \cite{gneiting2007strictly, ovadia2019can, salinas2020deepar}. We therefore adopt four complementary event-time and event-type metrics covering temporal accuracy, discrimination, probabilistic fit, and calibration: (1) Expected Event Time Error (EETE), the absolute error between ground-truth and expected predicted event times (7-day reference); (2) Event-Type Macro-F1 (ET-F1), balanced classification across valid transition types (e.g., $S \rightarrow E, E \rightarrow I, I \rightarrow R/D$); (3) Joint Event-Time Negative Log-Likelihood (NLL), penalizing low probability mass on realized outcomes; and (4) Event-Time Joint Brier Score (Brier), measuring calibration and sharpness. Metrics are computed on event-bearing states and averaged over rolling windows; lower is better for EETE, NLL, and Brier, and higher is better for ET-F1.

\subsection{Results}
\label{sec:exp-results}

\begin{figure*}[h]
    \centering
    \includegraphics[width=\textwidth]{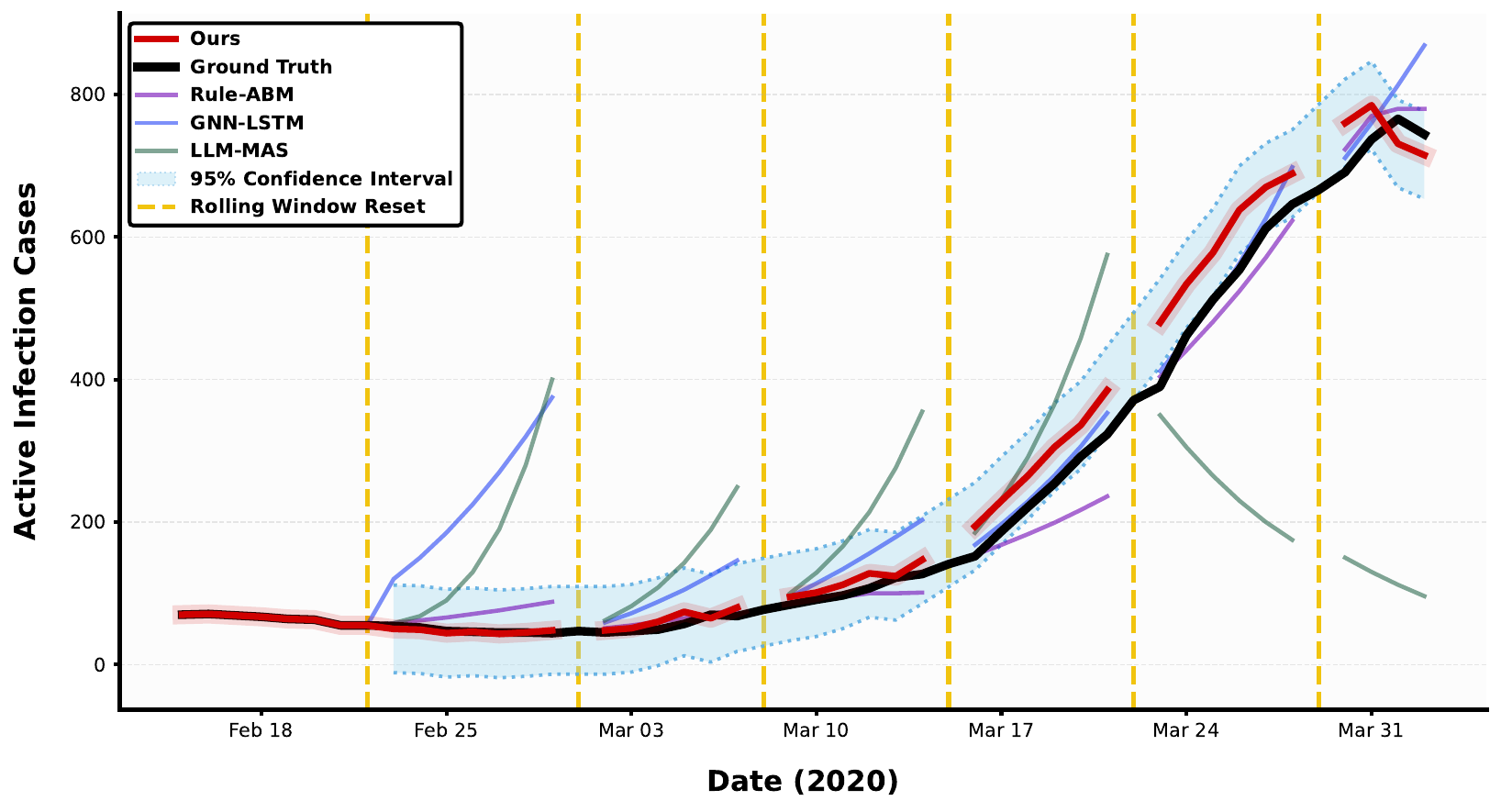}
   \caption{
\textbf{SEIRD dynamics under rolling-window forecasting.}
Infection trajectories for a 1{,}000-agent COVID-19 simulation comparing Rule-ABM, neural and LLM baselines with our model.
}
    \label{fig:seird}
\end{figure*}

\subsubsection{ANCHOR Cluster Interpretation}

\texttt{ANCHOR} identifies four behavioral control regimes in the Singapore COVID-19 simulation of 1,000 agents, defined by how agents reconfigure behavior across home, work, and community contexts. Figure ~\ref{fig:anchor-intepretation} reports normalized scores of the five dominant contextual motifs per cluster. Natural-language motif interpretations for the shown 5 dominant motifs are provided in Appendix Table~\ref{tab:anchor-motif-interpretation}. Complementarily, cross-contextual behavioral fingerprints are shown in Appendix fig. \ref{fig:anchor-radial-plot}. Stable Responders (Cluster~1) exhibit near-zero motif magnitudes with low variance, indicating smooth proportional adaptation and forming the behavioral inertia that largely determines the mean $S \rightarrow E$ trajectory. Environment-Driven Responders (Cluster~2) show weak intrinsic control and compressed motif distributions, reacting strongly to external contexts (e.g., workplace or community restrictions) with limited proactive mitigation. Self-Regulating Responders (Cluster~3) display strong, coherent motifs, actively suppressing contacts and increasing isolation and mask usage when transitioning between private and public contexts; despite their smaller population share, they disproportionately modulate $S \rightarrow E$ transitions through context-aware control. Volatile Responders (Cluster~4) exhibit wide, mixed-sign motif distributions with high variance, reflecting unstable control policies in which small contextual shifts induce large, inconsistent behavioral changes and act as primary sources of epidemic uncertainty.
None of the clusters is separable by age, static behavior, or contact structure alone. Instead, \texttt{ANCHOR} identifies functional equivalence classes defined by contextual control logic, yielding a mechanistic view of population dynamics beyond demographics or network position. See additional results and interpretations in Appendix Sec. \ref{supp-sec:anchor}.

\subsubsection{Quantitative Evaluation}
Table~\ref{tab:eval-quant-general} reports state-transition performance across epidemiology, finance, and social diffusion. \textit{PhysicsAgentABM} achieves the best results across all four metrics: EETE, ET-F1, NLL, and Brier in every domain, indicating superior temporal accuracy, event discrimination, and calibration. Mechanistic baselines underperform due to rigid rules under non-stationarity, while pure neural and LLM-based models attain competitive ET-F1 but exhibit higher temporal error and weaker calibration. Hybrid approaches without explicit uncertainty modeling show unstable performance. Overall, \texttt{PhysicsAgentABM} delivers robust, well-calibrated performance across domains, highlighting the benefits of hierarchical abstraction and uncertainty-aware fusion.

\subsubsection{Qualitative Evaluation}

\paragraph{COVID-19 Dynamics:} 

Figure \ref{fig:seird} illustrates epidemiological forecasting under rolling-window evaluation. PhysicsAgentABM closely tracks the ground-truth epidemic trajectory across repeated train–test resets, capturing pre-peak acceleration, aligning with the late-March infection peak, and rapidly adapting to the post-Circuit Breaker decline. In contrast, Rule-ABM fails to reproduce peak sharpness due to its inability to model behavioral heterogeneity and network effects, GNN-LSTM overshoots recovery and exhibits miscalibrated uncertainty, and LLM-MAS shows improved peak alignment but degrades over longer horizons due to temporal inconsistency. These improvements arise from a hierarchical division of labor across clustered agents. Tool-grounded symbolic reasoning enforces policy and contact-driven constraints during regime shifts, while neural predictors capture intra-cluster heterogeneity and temporal momentum during steady transmission. Epistemic fusion adaptively reweights these signals over rolling windows, enabling stable yet responsive forecasts.

\paragraph{Market Sentiment Diffusion:} 

\begin{figure*}[h]
    \centering
    \includegraphics[width=\textwidth]{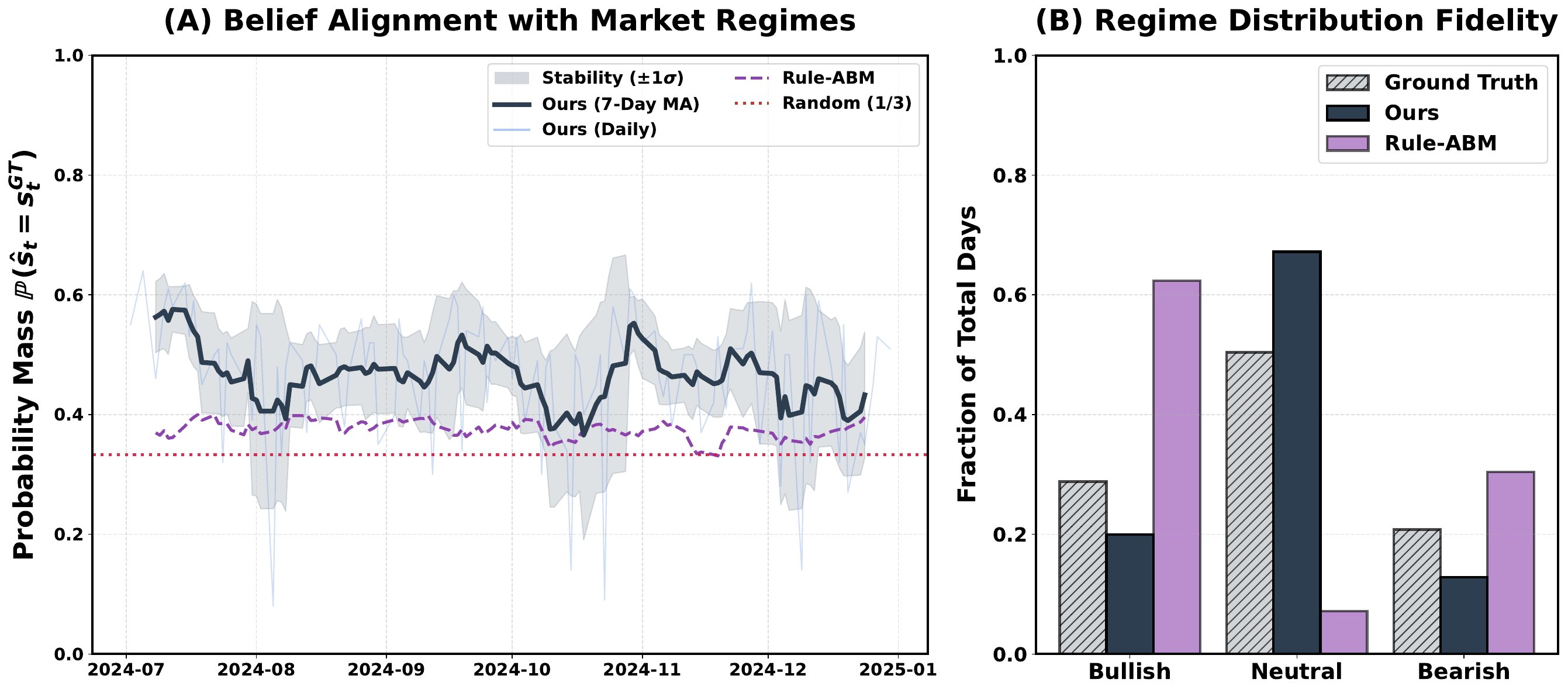}
    \caption{
\textbf{Population-level market belief dynamics inferred by the agent ecosystem.}
\textbf{(A)} Probability mass assigned to the realized market sentiment regime (S\&P 500) over time. Daily belief estimates fluctuate with market volatility(light blue line), while the 7-day moving average reveals a stable, coherent belief trajectory. Compared to a rule-based ABM, our model maintains higher and more persistent alignment with realized regimes, indicating superior collective belief formation under uncertainty. \textbf{(B)} Our model recovers a balanced regime distribution consistent with the realized market structure, while the rule-based ABM exhibits regime bias.
}
\label{fig:finance-qual}
\end{figure*}

We evaluate whether our ecosystem of agents can form coherent, uncertainty-aware beliefs about investment market regimes from weak, indirect signals, thereby developing an intuitive basis for anticipating market direction. Because individual sentiment is unobservable, inference relies on population-level daily S\&P 500 index values as a macro proxy, aligning agent states with realized market regimes (e.g., Bullish/Bearish). Under this framing, \texttt{PhysicsAgentABM} outperforms all baselines in temporal alignment, regime detection, and calibration (Table~\ref{tab:eval-quant-general}). Figure~\ref{fig:finance-qual}A shows that the model assigns elevated but realistic probability mass (60–75\%) to the realized daily regime, exceeding a random baseline while correctly reducing confidence during volatile periods (e.g., the October~2024 VIX spike). Despite short-term fluctuations, weekly averages remain smooth and aligned, indicating stable population-level belief formation. Figure~\ref{fig:finance-qual}B further shows close agreement with the empirical regime distribution, capturing neutral-dominant phases and sustained bearish shifts during drawdowns, whereas rule-based ABMs overrepresent bearish regimes and neural models exhibit regime imbalance. These gains arise from decoupling signal volatility from belief formation: symbolic reasoning interprets macro events, neural predictors capture temporal trends, and epistemic fusion suppresses transient noise.

\paragraph{Attention Lifecycle:} 

\begin{figure}[h]
    \centering
    \includegraphics[width=\columnwidth]{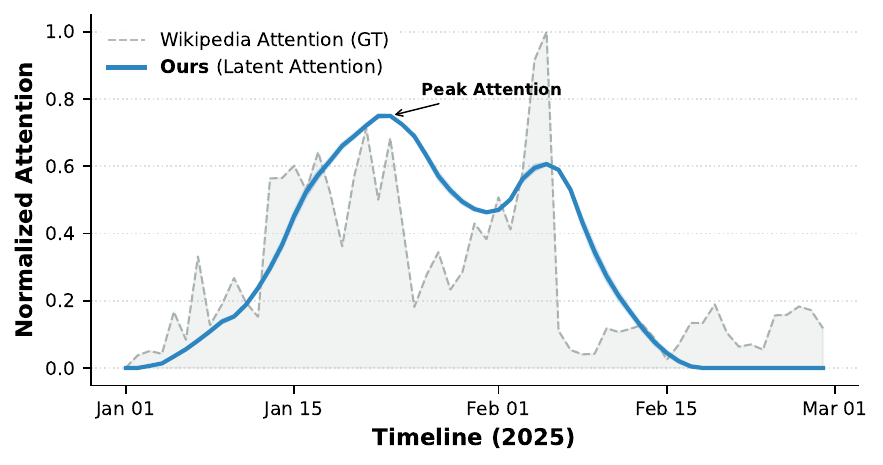}
    \caption{
    \textbf{Attention plot of interested people.}
    We evaluate attention diffusion using a rolling-window forecasting protocol, training on 28 days of historical data and predicting attention over the subsequent 7 days. Ground-truth attention is obtained from normalized daily Wikipedia pageviews, while model outputs are continuous probabilistic attention estimates.
    }
    \label{fig:attention-plot}
\end{figure}

We analyze the US population-level attention to a high-salience topic (\emph{Climate Change}) over a 90-day horizon (Dec~1,~2024 Feb~28,~2025), modeling attention dynamics through three latent states: \textbf{\emph{Unaware}, \emph{Interested}, and \emph{Fatigued}}. Because individual-level attention states are unobservable, we use normalized daily Wikipedia pageviews (attention index) as a noisy macro-level observation signal. As shown in Fig.~\ref{fig:attention-plot}, the observed pageview trajectory exhibits sharp, high-variance fluctuations driven by discrete exogenous events; most notably the LA wildfires in early January and COP28-related announcements in late January, followed by rapid decay as media salience subsides.

Using a small population of 250 agents, \textsc{PhysicsAgentABM} infers a smooth latent attention trajectory that integrates these observations over time, capturing the underlying evolution of public attention span rather than reacting to individual spikes. The inferred trajectory follows a characteristic \textbf{S-shaped diffusion pattern} \cite{orr2003diffusion,qiu2018deepinf}, with gradual emergence in early December, sustained amplification through mid-to-late January as repeated events reinforce salience, and a fatigue-driven decline through February. This behavior reflects the modeling objective of recovering coherent population attention regimes instead of reproducing platform-level volatility that conflates transient exposure with persistent engagement. The inferred peak aligns temporally with the window of elevated activity while remaining stable in the face of short-lived perturbations, and the post-peak phase exhibits smooth disengagement rather than erratic drop-off. Quantitatively, this regime-level inference achieves strong temporal alignment and calibration in social diffusion, yielding lower EETE (2.48) and Brier score (0.12) than all evaluated baselines (Table~\ref{tab:eval-quant-general}).

\subsection{Case Study: Singapore COVID-19 Circuit Breaker}

\begin{figure}[h]
    \centering
    \includegraphics[width=\columnwidth]{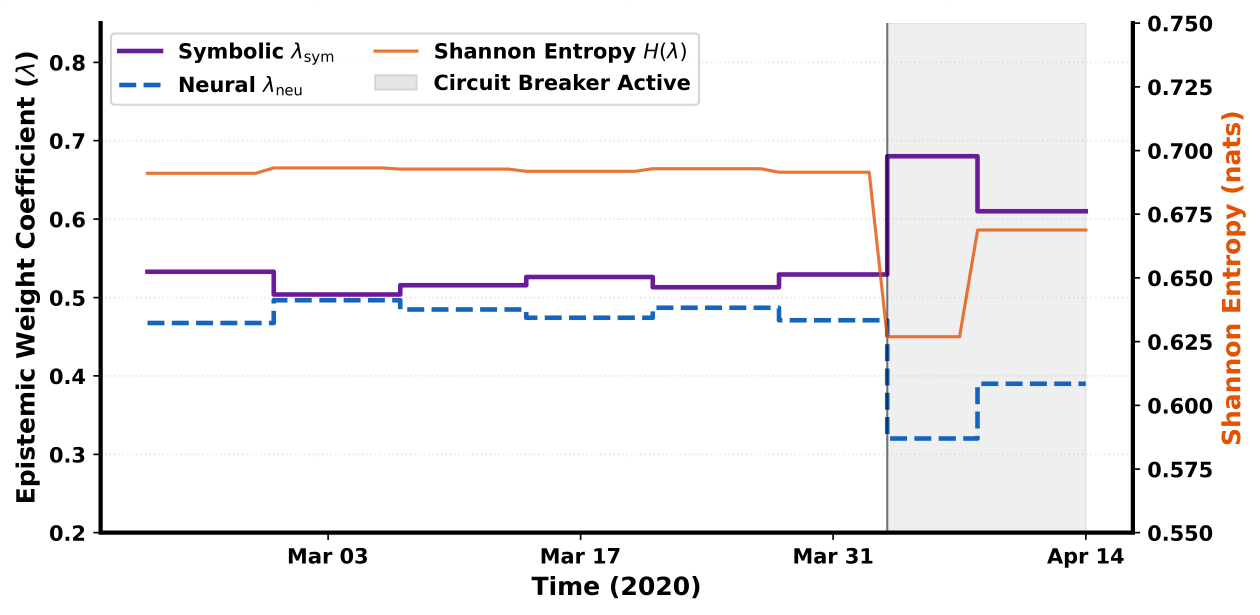}
  \caption{
\textbf{Epistemic fusion dynamics during the Singapore Circuit Breaker.}
At policy onset (April~3), symbolic inference is upweighted with a transient entropy spike, followed by rebalancing as dynamics stabilize.
}
    \label{fig:epistemic-weight-tracking}
\end{figure}

We evaluate \texttt{PhysicsAgentABM} under Singapore’s Circuit Breaker, an abrupt policy shock announced April~3, enforced April~7, 2020, using $N{=}1000$ agents. Before intervention, the model tracks exponential growth and predicts the infection peak within three days of the observed maximum (Fig.~\ref{fig:seird-recovery}). At policy onset, $S \rightarrow E$ transitions are sharply suppressed, infections decline, and recoveries accelerate, closely matching ground truth, while neural and LLM baselines overshoot due to temporal inertia. \texttt{PhysicsAgentABM} preserves micro-heterogeneity under lockdown, capturing residual dormitory transmission and intensified household spread. As shown in Fig.~\ref{fig:epistemic-weight-tracking}, epistemic fusion rapidly upweights symbolic inference ($\lambda_{\text{sym}}$) at the shock, enabling immediate incorporation of exogenous constraints and a controlled rise in predictive entropy, reflecting regime uncertainty. It then rebalances toward neural prediction as dynamics stabilize, yielding calibrated uncertainty contraction and linear recovery. This adaptive rebalancing produces immediate trajectory reversal and strong calibration gains (Brier $=0.16$, $76\%$ reduction vs.\ LLM baselines) with improved temporal alignment and event detection (Table~\ref{tab:eval-quant-general}). See Appendix Sec. \ref{supp-sec:covid-case-study} for additional discussion.



\section{Cost and Scalability Analysis}
\label{sec:cost-analysis}

We analyze the cost and scalability of \textsc{PhysicsAgentABM} in terms of token usage, API calls, and wall-clock time on a single A100 GPU with 50 asynchronous API calls (Table~\ref{tab:cost_scaling}). Unlike flat LLM-based GABMs that invoke high-context reasoning independently for every agent at every timestep, \textsc{PhysicsAgentABM} centralizes symbolic inference at the cluster level and sparsifies agent-level LLM usage, substantially reducing cost and latency while preserving heterogeneity and predictive fidelity. Experiments are conducted on epidemiological dynamics with $N{=}1000$ agents partitioned into $M{=}4$ clusters, each coordinated by 5 \textsc{StateAgents} and one \textsc{MetaAgent}. Let $\alpha \in (0,1]$ denote the fraction of agents invoking lightweight LLM reasoning per timestep (e.g., for rare or ambiguous transitions such as $S \rightarrow E$); all other agents realize transitions via fused cluster-level hazards, neighborhood states from the adjacency matrix $A$, and local short-term memory. At most, an agent issues one additional LLM call for text generation.

Relative to a flat GABM baseline, API calls drop from 8{,}250 to 1{,}233 ($6.7\times$). Total token usage decreases from 2.3M tokens per timestep to 0.79M at $\alpha{=}1.0$ and 0.49M at $\alpha{=}0.6$, corresponding to $2.9\times$ and $4.7\times$ reductions. Under GPT-4o-mini pricing, per-timestep cost falls from \$0.48 to \$0.23 and \$0.14 ($2.1$--$3.4\times$ savings). Wall-clock runtime similarly decreases from $\sim$300\,s to $\sim$40\,s and $\sim$24\,s per timestep, yielding $7.5$--$12.5\times$ speedups. Predictive performance is unaffected, with ET-F1 fixed at 0.81 across all $\alpha$ values.

\begin{table}[h]
\centering
\caption{
\textbf{Per-timestep cost, performance and scalability metrics} comparing a flat GABM baseline and \textsc{PhysicsAgentABM} under varying $\alpha$ .
}
\label{tab:cost_scaling}

\small
\setlength{\tabcolsep}{3pt}
\renewcommand{\arraystretch}{1.1}
\begin{tabular}{l | c | c | c | c}
\toprule
\rowcolor{headercolor} \textbf{Metric} & \textbf{Flat GABM} & \multicolumn{3}{c}{\textbf{PhysicsAgentABM}} \\
\rowcolor{headercolor} \textbf{(/timestep)} & & $\boldsymbol{\alpha{=}1.0}$ & $\boldsymbol{\alpha{=}0.75}$ & $\boldsymbol{\alpha{=}0.6}$ \\
\midrule
API calls & 8,250 & 1,083 & 833 & 683 \\
\rowcolor{rowcolor} Prompt tokens & 2,000,000 & 540,000 & 415,000 & 340,000 \\
Completion tokens & 300,000 & 250,000 & 187,500 & 150,000 \\
\rowcolor{rowcolor} Total tokens & 2,300,000 & 790,000 & 602,500 & 490,000 \\
WCT (s) & 300 & 40 & 30 & 24 \\
\rowcolor{rowcolor} Total WCT (mins) & 500 & 66 & 50 & 40 \\
\textbf{ET-F1} & \textbf{0.62} & \textbf{0.81} & \textbf{0.81} & \textbf{0.81} \\
\bottomrule
\end{tabular}
\end{table}

\begin{figure}[h]
    \centering
    \includegraphics[width=\columnwidth]{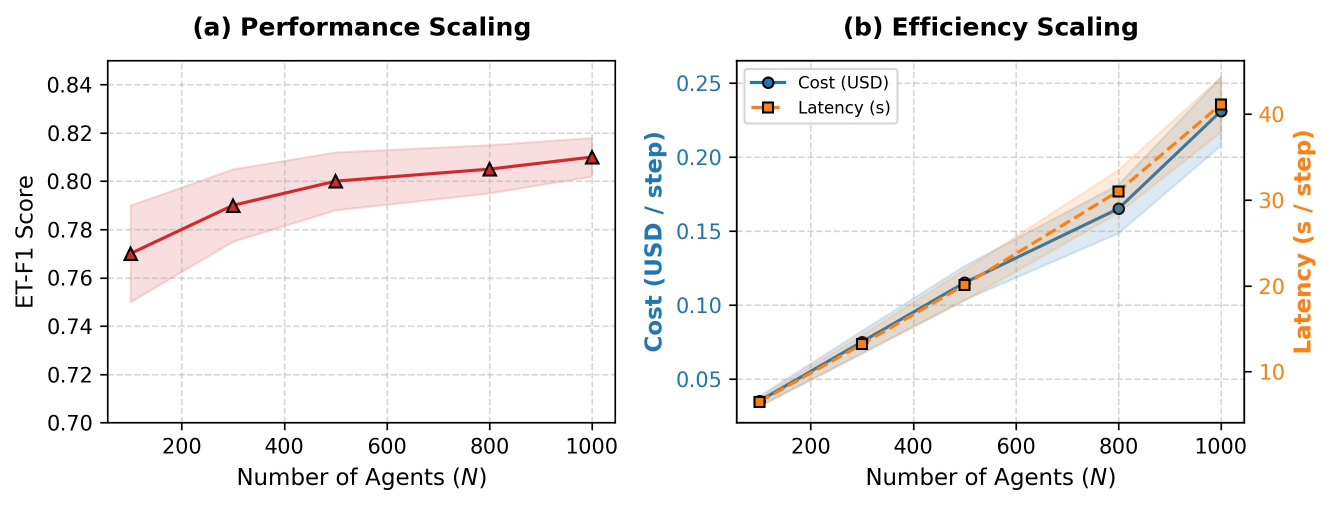}
    \caption{
\textbf{Cost, latency, and performance scaling with population size.}
(a) ET-F1 remains stable or improves slightly as the number of agents increases, reflecting more reliable cluster-level transition performance without loss of behavioral heterogeneity.
(b) Per-timestep cost and latency scale near linearly with $N$, as symbolic reasoning is amortized at the cluster level while agent-level realization remains lightweight and parallelizable.
}
    \label{fig:cost-latency-scaling}
\end{figure}

Figure~\ref{fig:cost-latency-scaling} shows scaling behavior with increasing population size. ET-F1 remains stable or improves slightly as $N$ grows, reflecting more reliable cluster-level statistics without over-smoothing or loss of behavioral diversity. Per-timestep cost and latency scale near linearly with $N$, confirming that symbolic reasoning cost is governed by the number of clusters rather than agents.
These results validate the central design principle of \textsc{PhysicsAgentABM}: expensive reasoning is amortized at the population level, while individual realization remains lightweight and parallelizable.

\subsection{Ablation}

\paragraph{PhysicsAgentABM Architecture:}
We ablate architectural components on the largest epidemiological setting (1{,}000 agents) under a fixed protocol. Removing clustering collapses population-level coordination, forcing agent-level reasoning that causes severe temporal misalignment and prohibitive latency (Table~\ref{tab:architecture-ablation}). Disabling epistemic fusion reveals complementary failures: neural-only variants are overconfident and delayed, while symbolic-only variants preserve structure but break under regime shifts. Within the symbolic pathway, removing state specialization or meta-coordination degrades performance at similar efficiency, confirming the necessity of role separation. With all components enabled, \texttt{PhysicsAgentABM} achieves calibrated, temporally aligned transitions with near-linear performance–efficiency scaling via cluster-level inference (See Appendix section ~\ref{supp-sec:calibration}). 

\paragraph{ANCHOR Component Analysis:}

\begin{figure}[t]
    \centering
    \includegraphics[width=\linewidth]{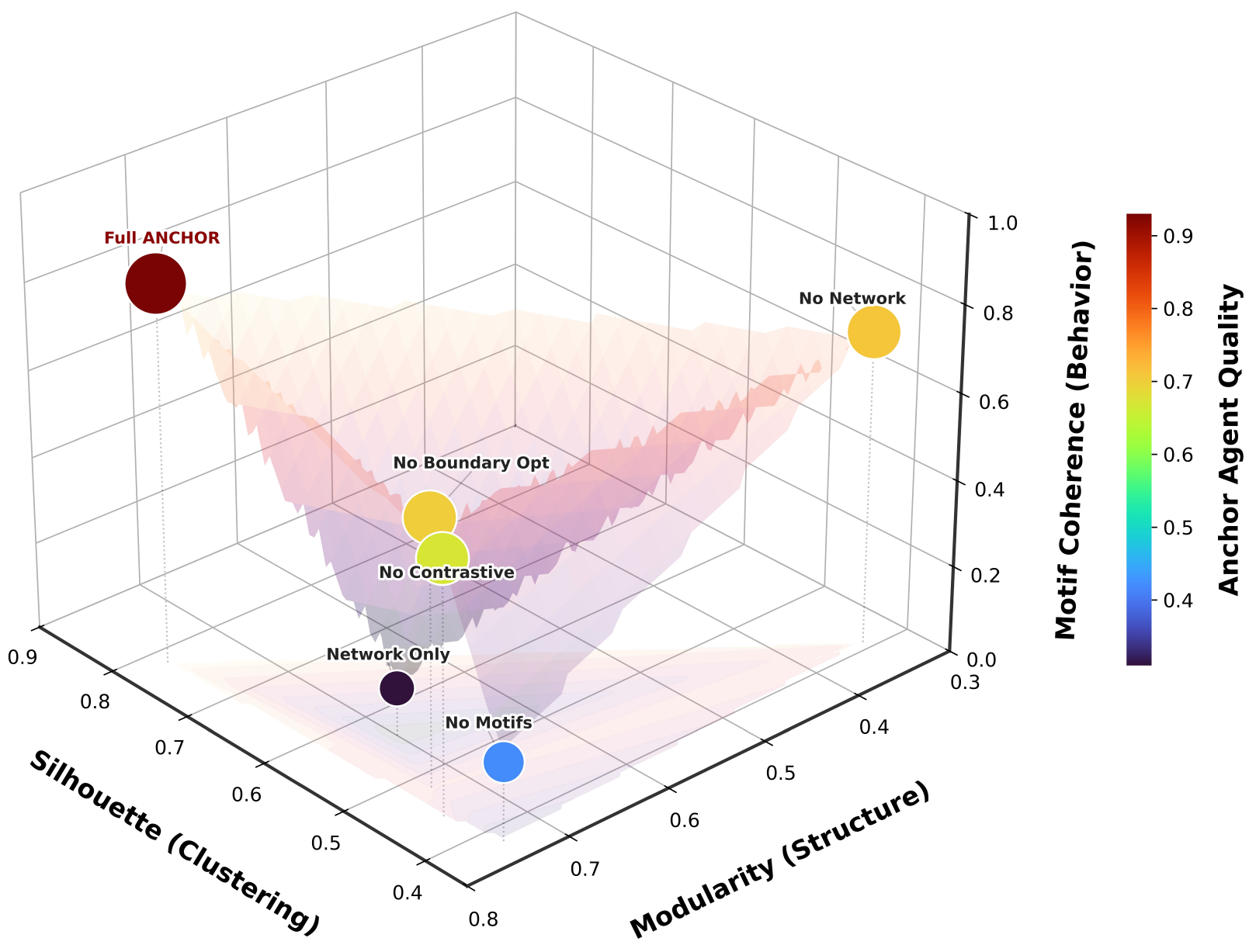}
    \caption{
    \textbf{Multi-objective ablation of \texttt{ANCHOR}.}
    Each point represents an ablation configuration evaluated by clustering separability (silhouette), structural coherence (modularity), and behavioral motif coherence.
    Marker size and color encode anchor-agent representativeness.
    The full \texttt{ANCHOR} model lies on the Pareto frontier, while variants removing motifs, contrastive alignment, or boundary optimization are systematically dominated.
    }
    \label{fig:anchor_ablation}
\end{figure}

We ablate \texttt{ANCHOR} under the same epidemiological setting to isolate failure modes in abstraction quality (Fig.~\ref{fig:anchor_ablation}). Removing behavioral motifs or contrastive cross-context alignment collapses clusters toward configuration-driven centroids, yielding high apparent structural separability but substantially degraded behavioral coherence. Structure-only variants preserve modularity yet fail to group agents by functional response, producing brittle abstractions that break under context shifts. Disabling boundary optimization further fragments clusters near regime transitions, indicating sensitivity to ambiguous state boundaries. In contrast, the full \texttt{ANCHOR} model is Pareto-dominant, jointly optimizing clustering separability, structural coherence, and contextual behavioral similarity.

Furthermore, \texttt{ANCHOR} separates agents based on context-dependent behavioral response patterns rather than static structures. As shown in Appendix Fig. \ref{fig:anchor-radial-plot}, clusters found by \texttt{ANCHOR} exhibit internally consistent yet nonlinearly reweighted behavior across home, work, and community contexts, enabling stable abstraction under policy and regime shifts.

\begin{table}[t]
\centering
\caption{PhysicsAgentABM architecture ablation.}
\label{tab:architecture-ablation}
\small
\setlength{\tabcolsep}{4pt}
\renewcommand{\arraystretch}{1.15}
\definecolor{headerorange}{RGB}{255, 235, 205} 

\rowcolors{2}{white}{gray!10}

\begin{tabular}{lccccc}
\toprule
\rowcolor{headerorange} 
\textbf{Configuration} & \makecell{\textbf{EETE} \\ $\downarrow$} & \makecell{\textbf{F1} \\ $\uparrow$} & \makecell{\textbf{NLL} \\ $\downarrow$} & \makecell{\textbf{Brier} \\ $\downarrow$} & \makecell{\textbf{Latency} \\ (s/step) $\downarrow$} \\
\midrule

\multicolumn{6}{l}{\textbf{Fusion}} \\
Neural-only   
& 3.15 & 0.62 & 0.96 & 0.31 & 0.6 \\
Symbolic-only 
& 2.98 & 0.66 & 0.88 & 0.27 & 35.8 \\
Naive fusion  
& 2.71 & 0.69 & 0.82 & 0.25 & 36.1 \\
\midrule

\multicolumn{6}{l}{\textbf{No clustering}} \\
Flat agents   
& 4.92 & 0.63 & 1.21 & 0.39 & 283 \\
\midrule

\multicolumn{6}{l}{\textbf{Symbolic Team}} \\
No state-agent 
& 2.64 & 0.61 & 0.79 & 0.23 & 268.4 \\
No meta-agent  
& 2.81 & 0.68 & 0.83 & 0.26 & 135.2 \\
\midrule

\rowcolor{white}
\textbf{PhysicsAgentABM} 
& \textbf{1.92} & \textbf{0.81} & \textbf{0.73} & \textbf{0.16} & \textbf{41.1} \\
\bottomrule
\end{tabular}
\end{table}

\section{Conclusion}
\label{sec:conclusion}
We introduce \textbf{PhysicsAgentABM}, a principled framework that shifts generative ABM from isolated agent simulation to population-level inference, together with \textbf{ANCHOR}, a contextual behavior-driven clustering mechanism guided by an LLM agent. Across diverse domains, PhysicsAgentABM consistently yields calibrated, temporally aligned dynamics while preserving heterogeneous individual realization. We believe this work establishes a system-level foundation for generative simulation
under real-world complexity.



\clearpage

\bibliography{main}

@article{wang2024survey,
  title={A survey on large language model based autonomous agents},
  author={Wang, Lei and Ma, Chen and Feng, Xueyang and Zhang, Zeyu and Yang, Hao and Zhang, Jingsen and Chen, Zhiyuan and Tang, Jiakai and Chen, Xu and Lin, Yankai and others},
  journal={Frontiers of Computer Science},
  volume={18},
  number={6},
  pages={186345},
  year={2024},
  publisher={Springer}
}

@article{orr2003diffusion,
  title={Diffusion of innovations, by Everett Rogers (1995)},
  author={Orr, Greg},
  journal={Retrieved January},
  volume={21},
  pages={2005},
  year={2003}
}

@article{andrews1995survey,
  title={Survey and critique of techniques for extracting rules from trained artificial neural networks},
  author={Andrews, Robert and Diederich, Joachim and Tickle, Alan B},
  journal={Knowledge-based systems},
  volume={8},
  number={6},
  pages={373--389},
  year={1995},
  publisher={Elsevier}
}

@inproceedings{park2023generative,
  title={Generative agents: Interactive simulacra of human behavior},
  author={Park, Joon Sung and O'Brien, Joseph and Cai, Carrie Jun and Morris, Meredith Ringel and Liang, Percy and Bernstein, Michael S},
  booktitle={Proceedings of the 36th annual acm symposium on user interface software and technology},
  pages={1--22},
  year={2023}
}

@article{manhaeve2018deepproblog,
  title={Deepproblog: Neural probabilistic logic programming},
  author={Manhaeve, Robin and Dumancic, Sebastijan and Kimmig, Angelika and Demeester, Thomas and De Raedt, Luc},
  journal={Advances in neural information processing systems},
  volume={31},
  year={2018}
}

@article{rossi2020temporal,
  title={Temporal graph networks for deep learning on dynamic graphs},
  author={Rossi, Emanuele and Chamberlain, Ben and Frasca, Fabrizio and Eynard, Davide and Monti, Federico and Bronstein, Michael},
  journal={arXiv preprint arXiv:2006.10637},
  year={2020}
}

@article{scarselli2008graph,
  title={The graph neural network model},
  author={Scarselli, Franco and Gori, Marco and Tsoi, Ah Chung and Hagenbuchner, Markus and Monfardini, Gabriele},
  journal={IEEE transactions on neural networks},
  volume={20},
  number={1},
  pages={61--80},
  year={2008},
  publisher={IEEE}
}

@article{hochreiter1997long,
  title={Long short-term memory},
  author={Hochreiter, Sepp and Schmidhuber, J{\"u}rgen},
  journal={Neural computation},
  volume={9},
  number={8},
  pages={1735--1780},
  year={1997},
  publisher={MIT press}
}

@article{terHoeven2025Mesa3,
  title = {Mesa 3: Agent-based modeling with Python in 2025},
  author = {ter Hoeven, Ewout and Kwakkel, Jan and Hess, Vincent and Pike, Thomas and Wang, Boyu and rht and Kazil, Jackie},
  journal = {Journal of Open Source Software},
  volume = {10},
  number = {107},
  pages = {7668},
  year = {2025},
  doi = {10.21105/joss.07668},
  url = {https://doi.org/10.21105/joss.07668}
}

@article{mckean1966class,
  title={A class of Markov processes associated with nonlinear parabolic equations},
  author={McKean Jr, Henry P},
  journal={Proceedings of the National Academy of Sciences},
  volume={56},
  number={6},
  pages={1907--1911},
  year={1966}
}

@article{owid-coronavirus,
    author = {Edouard Mathieu and Hannah Ritchie and Lucas Rodés-Guirao and Cameron Appel and Daniel Gavrilov and Charlie Giattino and Joe Hasell and Bobbie Macdonald and Saloni Dattani and Diana Beltekian and Esteban Ortiz-Ospina and Max Roser},
    title = {COVID-19 Pandemic},
    journal = {Our World in Data},
    year = {2020},
    note = {https://ourworldindata.org/coronavirus}
}

@inproceedings{guo2024large,
  title={Large Language Model Based Multi-agents: A Survey of Progress and Challenges},
  author={Guo, Taicheng and Chen, Xiuying and Wang, Yaqi and Chang, Ruidi and Pei, Shichao and Chawla, Nitesh V and Wiest, Olaf and Zhang, Xiangliang},
  booktitle={IJCAI},
  year={2024}
}

@article{bonabeau2002agent,
  title={Agent-based modeling: Methods and techniques for simulating human systems},
  author={Bonabeau, Eric},
  journal={Proceedings of the national academy of sciences},
  volume={99},
  number={suppl\_3},
  pages={7280--7287},
  year={2002},
  publisher={National Academy of Sciences}
}

@article{hethcote2000mathematics,
  title={The mathematics of infectious diseases},
  author={Hethcote, Herbert W},
  journal={SIAM review},
  volume={42},
  number={4},
  pages={599--653},
  year={2000},
  publisher={SIAM}
}

@book{hurd2016contagion,
  title={Contagion! Systemic risk in financial networks},
  author={Hurd, Thomas R and others},
  volume={42},
  year={2016},
  publisher={Springer}
}

@article{watts2002simple,
  title={A simple model of global cascades on random networks},
  author={Watts, Duncan J},
  journal={Proceedings of the National Academy of Sciences},
  volume={99},
  number={9},
  pages={5766--5771},
  year={2002},
  publisher={The National Academy of Sciences}
}

@article{gillespie1977exact,
  title={Exact stochastic simulation of coupled chemical reactions},
  author={Gillespie, Daniel T},
  journal={The journal of physical chemistry},
  volume={81},
  number={25},
  pages={2340--2361},
  year={1977},
  publisher={ACS Publications}
}

@article{pastor2015epidemic,
  title={Epidemic processes in complex networks},
  author={Pastor-Satorras, Romualdo and Castellano, Claudio and Van Mieghem, Piet and Vespignani, Alessandro},
  journal={Reviews of modern physics},
  volume={87},
  number={3},
  pages={925--979},
  year={2015},
  publisher={APS}
}

@article{granovetter1978threshold,
  title={Threshold models of collective behavior},
  author={Granovetter, Mark},
  journal={American journal of sociology},
  volume={83},
  number={6},
  pages={1420--1443},
  year={1978},
  publisher={University of Chicago Press}
}

@incollection{epstein2012generative,
  title={Generative social science: Studies in agent-based computational modeling},
  author={Epstein, Joshua M},
  booktitle={Generative Social Science},
  year={2012},
  publisher={Princeton University Press}
}

@inproceedings{hong2023metagpt,
  title={MetaGPT: Meta programming for a multi-agent collaborative framework},
  author={Hong, Sirui and Zhuge, Mingchen and Chen, Jonathan and Zheng, Xiawu and Cheng, Yuheng and Wang, Jinlin and Zhang, Ceyao and Wang, Zili and Yau, Steven Ka Shing and Lin, Zijuan and others},
  booktitle={The twelfth international conference on learning representations},
  year={2023}
}

@inproceedings{qian2024chatdev,
  title={Chatdev: Communicative agents for software development},
  author={Qian, Chen and Liu, Wei and Liu, Hongzhang and Chen, Nuo and Dang, Yufan and Li, Jiahao and Yang, Cheng and Chen, Weize and Su, Yusheng and Cong, Xin and others},
  booktitle={Proceedings of the 62nd Annual Meeting of the Association for Computational Linguistics (Volume 1: Long Papers)},
  pages={15174--15186},
  year={2024}
}

@article{gao2024agentscope,
  title={Agentscope: A flexible yet robust multi-agent platform},
  author={Gao, Dawei and Li, Zitao and Pan, Xuchen and Kuang, Weirui and Ma, Zhijian and Qian, Bingchen and Wei, Fei and Zhang, Wenhao and Xie, Yuexiang and Chen, Daoyuan and others},
  journal={arXiv preprint arXiv:2402.14034},
  year={2024}
}

@inproceedings{du2023improving,
  title={Improving factuality and reasoning in language models through multiagent debate},
  author={Du, Yilun and Li, Shuang and Torralba, Antonio and Tenenbaum, Joshua B and Mordatch, Igor},
  booktitle={Forty-first International Conference on Machine Learning},
  year={2023}
}

@inproceedings{mou2024unveiling,
  title={Unveiling the truth and facilitating change: Towards agent-based large-scale social movement simulation},
  author={Mou, Xinyi and Wei, Zhongyu and Huang, Xuan-Jing},
  booktitle={Findings of the Association for Computational Linguistics: ACL 2024},
  pages={4789--4809},
  year={2024}
}

@article{gai2010contagion,
  title={Contagion in Financial Networks},
  author={Gai, Prasanna and Kapadia, Sujit},
  journal={Proceedings of the Royal Society A},
  year={2010}
}

@book{murphy2012machine,
  title={Machine Learning: A Probabilistic Perspective},
  author={Murphy, Kevin P.},
  publisher={MIT Press},
  year={2012}
}

@article{gneiting2007strictly,
  title={Strictly proper scoring rules, prediction, and estimation},
  author={Gneiting, Tilmann and Raftery, Adrian E},
  journal={Journal of the American statistical Association},
  volume={102},
  number={477},
  pages={359--378},
  year={2007},
  publisher={Taylor \& Francis}
}

@article{ovadia2019can,
  title={Can you trust your model's uncertainty? evaluating predictive uncertainty under dataset shift},
  author={Ovadia, Yaniv and Fertig, Emily and Ren, Jie and Nado, Zachary and Sculley, David and Nowozin, Sebastian and Dillon, Joshua and Lakshminarayanan, Balaji and Snoek, Jasper},
  journal={Advances in neural information processing systems},
  volume={32},
  year={2019}
}

@article{salinas2020deepar,
  title={DeepAR: Probabilistic forecasting with autoregressive recurrent networks},
  author={Salinas, David and Flunkert, Valentin and Gasthaus, Jan and Januschowski, Tim},
  journal={International journal of forecasting},
  volume={36},
  number={3},
  pages={1181--1191},
  year={2020},
  publisher={Elsevier}
}

@article{krishnan2017structured,
  title={Structured Inference Networks for Nonlinear State Space Models},
  author={Krishnan, Rahul and others},
  journal={AAAI},
  year={2017}
}

@article{kipf2017semi,
  title={Semi-Supervised Classification with Graph Convolutional Networks},
  author={Kipf, Thomas N. and Welling, Max},
  journal={ICLR},
  year={2017}
}

@article{hamilton2017inductive,
  title={Inductive Representation Learning on Large Graphs},
  author={Hamilton, William L. and others},
  journal={NeurIPS},
  year={2017}
}

@inproceedings{xiang2022temporal,
  title={Temporal and heterogeneous graph neural network for financial time series prediction},
  author={Xiang, Sheng and Cheng, Dawei and Shang, Chencheng and Zhang, Ying and Liang, Yuqi},
  booktitle={Proceedings of the 31st ACM international conference on information \& knowledge management},
  pages={3584--3593},
  year={2022}
}

@article{ying2018hierarchical,
  title={Hierarchical graph representation learning with differentiable pooling},
  author={Ying, Zhitao and You, Jiaxuan and Morris, Christopher and Ren, Xiang and Hamilton, Will and Leskovec, Jure},
  journal={Advances in neural information processing systems},
  volume={31},
  year={2018}
}

@article{lowe2017multi,
  title={Multi-agent actor-critic for mixed cooperative-competitive environments},
  author={Lowe, Ryan and Wu, Yi I and Tamar, Aviv and Harb, Jean and Pieter Abbeel, OpenAI and Mordatch, Igor},
  journal={Advances in neural information processing systems},
  volume={30},
  year={2017}
}

@article{battaglia2018relational,
  title={Relational inductive biases, deep learning, and graph networks},
  author={Battaglia, Peter W and Hamrick, Jessica B and Bapst, Victor and Sanchez-Gonzalez, Alvaro and Zambaldi, Vinicius and Malinowski, Mateusz and Tacchetti, Andrea and Raposo, David and Santoro, Adam and Faulkner, Ryan and others},
  journal={arXiv preprint arXiv:1806.01261},
  year={2018}
}

@article{velivckovic2017graph,
  title={Graph attention networks},
  author={Veli{\v{c}}kovi{\'c}, Petar and Cucurull, Guillem and Casanova, Arantxa and Romero, Adriana and Lio, Pietro and Bengio, Yoshua},
  journal={arXiv preprint arXiv:1710.10903},
  year={2017}
}

@inproceedings{fang2021spatial,
  title={Spatial-temporal graph ode networks for traffic flow forecasting},
  author={Fang, Zheng and Long, Qingqing and Song, Guojie and Xie, Kunqing},
  booktitle={Proceedings of the 27th ACM SIGKDD conference on knowledge discovery \& data mining},
  pages={364--373},
  year={2021}
}

@inproceedings{kipf2018neural,
  title={Neural relational inference for interacting systems},
  author={Kipf, Thomas and Fetaya, Ethan and Wang, Kuan-Chieh and Welling, Max and Zemel, Richard},
  booktitle={International conference on machine learning},
  pages={2688--2697},
  year={2018},
  organization={Pmlr}
}

@inproceedings{deng2020cola,
  title={Cola-GNN: Cross-location attention based graph neural networks for long-term ILI prediction},
  author={Deng, Songgaojun and Wang, Shusen and Rangwala, Huzefa and Wang, Lijing and Ning, Yue},
  booktitle={Proceedings of the 29th ACM international conference on information \& knowledge management},
  pages={245--254},
  year={2020}
}

@article{xu2021hist,
  title={Hist: A graph-based framework for stock trend forecasting via mining concept-oriented shared information},
  author={Xu, Wentao and Liu, Weiqing and Wang, Lewen and Xia, Yingce and Bian, Jiang and Yin, Jian and Liu, Tie-Yan},
  journal={arXiv preprint arXiv:2110.13716},
  year={2021}
}

@book{centola2018behavior,
  title={How behavior spreads: The science of complex contagions},
  author={Centola, Damon},
  volume={3},
  year={2018},
  publisher={Princeton University Press Princeton, NJ}
}

@book{gelman1995bayesian,
  title={Bayesian data analysis},
  author={Gelman, Andrew and Carlin, John B and Stern, Hal S and Rubin, Donald B},
  year={1995},
  publisher={Chapman and Hall/CRC}
}

@article{diebold2014network,
  title={On the network topology of variance decompositions: Measuring the connectedness of financial firms},
  author={Diebold, Francis X and Y{\i}lmaz, Kamil},
  journal={Journal of econometrics},
  volume={182},
  number={1},
  pages={119--134},
  year={2014},
  publisher={Elsevier}
}

@article{billio2012econometric,
  title={Econometric measures of connectedness and systemic risk in the finance and insurance sectors},
  author={Billio, Monica and Getmansky, Mila and Lo, Andrew W and Pelizzon, Loriana},
  journal={Journal of financial economics},
  volume={104},
  number={3},
  pages={535--559},
  year={2012},
  publisher={Elsevier}
}

@article{block2020social,
  title={Social network-based distancing strategies to flatten the COVID-19 curve in a post-lockdown world},
  author={Block, Per and Hoffman, Marion and Raabe, Isabel J and Dowd, Jennifer Beam and Rahal, Charles and Kashyap, Ridhi and Mills, Melinda C},
  journal={Nature human behaviour},
  volume={4},
  number={6},
  pages={588--596},
  year={2020},
  publisher={Nature Publishing Group UK London}
}

@article{chang2021mobility,
  title={Mobility network models of COVID-19 explain inequities and inform reopening},
  author={Chang, Serina and Pierson, Emma and Koh, Pang Wei and Gerardin, Jaline and Redbird, Beth and Grusky, David and Leskovec, Jure},
  journal={Nature},
  volume={589},
  number={7840},
  pages={82--87},
  year={2021},
  publisher={Nature Publishing Group UK London}
}

@article{yi2018neural,
  title={Neural-symbolic vqa: Disentangling reasoning from vision and language understanding},
  author={Yi, Kexin and Wu, Jiajun and Gan, Chuang and Torralba, Antonio and Kohli, Pushmeet and Tenenbaum, Josh},
  journal={Advances in neural information processing systems},
  volume={31},
  year={2018}
}

@inproceedings{kempe2003maximizing,
  title={Maximizing the spread of influence through a social network},
  author={Kempe, David and Kleinberg, Jon and Tardos, {\'E}va},
  booktitle={Proceedings of the ninth ACM SIGKDD international conference on Knowledge discovery and data mining},
  pages={137--146},
  year={2003}
}

@inproceedings{qiu2018deepinf,
  title={Deepinf: Social influence prediction with deep learning},
  author={Qiu, Jiezhong and Tang, Jian and Ma, Hao and Dong, Yuxiao and Wang, Kuansan and Tang, Jie},
  booktitle={Proceedings of the 24th ACM SIGKDD international conference on knowledge discovery \& data mining},
  pages={2110--2119},
  year={2018}
}

@inproceedings{ye2024language,
  title={Language is all a graph needs},
  author={Ye, Ruosong and Zhang, Caiqi and Wang, Runhui and Xu, Shuyuan and Zhang, Yongfeng},
  booktitle={Findings of the association for computational linguistics: EACL 2024},
  pages={1955--1973},
  year={2024}
}

@article{pan2024unifying,
  title={Unifying large language models and knowledge graphs: A roadmap},
  author={Pan, Shirui and Luo, Linhao and Wang, Yufei and Chen, Chen and Wang, Jiapu and Wu, Xindong},
  journal={IEEE Transactions on Knowledge and Data Engineering},
  volume={36},
  number={7},
  pages={3580--3599},
  year={2024},
  publisher={IEEE}
}

@article{blondel2008fast,
  title={Fast unfolding of communities in large networks},
  author={Blondel, Vincent D and Guillaume, Jean-Loup and Lambiotte, Renaud and Lefebvre, Etienne},
  journal={Journal of statistical mechanics: theory and experiment},
  volume={2008},
  number={10},
  pages={P10008},
  year={2008},
  publisher={IOP Publishing}
}

@article{von2007tutorial,
  title={A tutorial on spectral clustering},
  author={Von Luxburg, Ulrike},
  journal={Statistics and computing},
  volume={17},
  number={4},
  pages={395--416},
  year={2007},
  publisher={Springer}
}

@article{newman2006modularity,
  title={Modularity and community structure in networks},
  author={Newman, Mark EJ},
  journal={Proceedings of the national academy of sciences},
  volume={103},
  number={23},
  pages={8577--8582},
  year={2006},
  publisher={National Academy of Sciences}
}

@article{wang2023plan,
  title={Plan-and-solve prompting: Improving zero-shot chain-of-thought reasoning by large language models},
  author={Wang, Lei and Xu, Wanyu and Lan, Yihuai and Hu, Zhiqiang and Lan, Yunshi and Lee, Roy Ka-Wei and Lim, Ee-Peng},
  journal={arXiv preprint arXiv:2305.04091},
  year={2023}
}

@article{zhu2023ghost,
  title={Ghost in the minecraft: Generally capable agents for open-world environments via large language models with text-based knowledge and memory},
  author={Zhu, Xizhou and Chen, Yuntao and Tian, Hao and Tao, Chenxin and Su, Weijie and Yang, Chenyu and Huang, Gao and Li, Bin and Lu, Lewei and Wang, Xiaogang and others},
  journal={arXiv preprint arXiv:2305.17144},
  year={2023}
}

@inproceedings{wu2024autogen,
  title={Autogen: Enabling next-gen LLM applications via multi-agent conversations},
  author={Wu, Qingyun and Bansal, Gagan and Zhang, Jieyu and Wu, Yiran and Li, Beibin and Zhu, Erkang and Jiang, Li and Zhang, Xiaoyun and Zhang, Shaokun and Liu, Jiale and others},
  booktitle={First Conference on Language Modeling},
  year={2024}
}

@article{li2023camel,
  title={Camel: Communicative agents for" mind" exploration of large language model society},
  author={Li, Guohao and Hammoud, Hasan and Itani, Hani and Khizbullin, Dmitrii and Ghanem, Bernard},
  journal={Advances in Neural Information Processing Systems},
  volume={36},
  pages={51991--52008},
  year={2023}
}

@inproceedings{yang2009combining,
  title={Combining link and content for community detection: a discriminative approach},
  author={Yang, Tianbao and Jin, Rong and Chi, Yun and Zhu, Shenghuo},
  booktitle={Proceedings of the 15th ACM SIGKDD international conference on Knowledge discovery and data mining},
  pages={927--936},
  year={2009}
}

@inproceedings{huang2022language,
  title={Language models as zero-shot planners: Extracting actionable knowledge for embodied agents},
  author={Huang, Wenlong and Abbeel, Pieter and Pathak, Deepak and Mordatch, Igor},
  booktitle={International conference on machine learning},
  pages={9118--9147},
  year={2022},
  organization={PMLR}
}

@article{hullermeier2021aleatoric,
  title={Aleatoric and epistemic uncertainty in machine learning: An introduction to concepts and methods},
  author={H{\"u}llermeier, Eyke and Waegeman, Willem},
  journal={Machine learning},
  volume={110},
  number={3},
  pages={457--506},
  year={2021},
  publisher={Springer}
}

@article{baltruvsaitis2018multimodal,
  title={Multimodal machine learning: A survey and taxonomy},
  author={Baltru{\v{s}}aitis, Tadas and Ahuja, Chaitanya and Morency, Louis-Philippe},
  journal={IEEE transactions on pattern analysis and machine intelligence},
  volume={41},
  number={2},
  pages={423--443},
  year={2018},
  publisher={IEEE}
}

@article{fortuin2019deep,
  title={Deep mean functions for meta-learning in gaussian processes},
  author={Fortuin, Vincent and R{\"a}tsch, Gunnar},
  journal={arXiv preprint arXiv:1901.08098},
  volume={8},
  year={2019}
}

@article{tran2017hierarchical,
  title={Hierarchical implicit models and likelihood-free variational inference},
  author={Tran, Dustin and Ranganath, Rajesh and Blei, David},
  journal={Advances in Neural Information Processing Systems},
  volume={30},
  year={2017}
}

@article{generative1000people2024,
  title={Generative Agent Simulations of 1,000 People},
  journal={arXiv},
  year={2024}
}

@article{socioverse2025,
  title={SocioVerse: Large-Scale Social Simulation with LLM Agents},
  journal={arXiv},
  year={2025}
}

@article{llmempowerabm2023,
  title={Large Language Models Empower Agent-Based Modeling},
  journal={arXiv},
  year={2023}
}

@article{abmgenai2024,
  title={Agent-Based Modelling Meets Generative AI in Social Network Simulations},
  journal={arXiv},
  year={2024}
}

@article{agentsociety2025,
  title={AgentSociety: Large-Scale Simulation of LLM-Driven Generative Agents},
  journal={arXiv},
  year={2025}
}

@article{llmtrust2024,
  title={Can Large Language Model Agents Simulate Human Trust Behavior?},
  journal={arXiv},
  year={2024}
}

@article{limitsagency2024,
  title={On the Limits of Agency in Agent-Based Models},
  journal={arXiv},
  year={2024}
}

@book{schaeffer2007graph,
  title={Graph Clustering},
  author={Schaeffer, Satu Elisa},
  publisher={Computer Science Review},
  year={2007}
}

@article{loukas2019graph,
  title={Graph Reduction via Spectral Sparsification},
  author={Loukas, Andreas},
  journal={JMLR},
  year={2019}
}
\bibliographystyle{icml2026}

\clearpage
\appendix
\clearpage
\onecolumn
\subsection*{Table of Contents}
\noindent

\textbf{A} \quad Additional Details on Singapore COVID-19 Case Study \dotfill \pageref{supp-sec:covid-case-study}

\textbf{B} \quad Additional Details on \texttt{ANCHOR} \dotfill \pageref{supp-sec:anchor}\\
\hspace*{2em} B.1 \quad Algorithm \dotfill \pageref{supp-sec:anchor-algo}\\
\hspace*{2em} B.2 \quad Additional Ablations \dotfill \pageref{suppl-sec:anchor-ablation}\\
\hspace*{2em} B.3 \quad Top 5 Dominant Motifs Description \dotfill \pageref{tab:anchor-motif-interpretation}

\textbf{C} \quad PhysicsAgentABM Calibration Analysis \dotfill \pageref{supp-sec:calibration}

\textbf{D} \quad Symbolic Reasoning Pathway \dotfill \pageref{supp-sec:symblic-reasoning}\\
\hspace*{2em} D.1 \quad MetaAgent Prompt \dotfill \pageref{box:meta-agent-prompt}\\
\hspace*{2em} D.2 \quad StateAgent Prompt \dotfill \pageref{box:state-agent-prompt}\\
\hspace*{2em} D.3 \quad EntityAgent Prompt \dotfill \pageref{box:entity-agent-prompt}\\
\hspace*{2em} D.4 \quad Symbolic Agentic Tools \dotfill \pageref{tab:symbolic_agents}

\textbf{E} \quad Neural Multimodal Pathway \dotfill \pageref{supp-sec:neural-pathway}\\
\hspace*{2em} E.1 \quad Details on Multimodal Data Across the Three Domains \dotfill \pageref{tab:neural_input_modalities}\\
\hspace*{2em} E.2 \quad Neural Model Architectural Details \dotfill \pageref{tab:neural-architecture}\\
\hspace*{2em} E.3 \quad Training Configuration and Parameters \dotfill \pageref{tab:nn-training-config}

\textbf{F} \quad Experimental Setup \& Design Principles \dotfill \pageref{supp-sec:dataset-construction}\\
\hspace*{2em} F.1 \quad Task Overview \dotfill \pageref{supp-sec:task-overview}\\
\hspace*{2em} F.2 \quad Dataset Overview \dotfill \pageref{supp-sec:dataset-overview}\\
\hspace*{2em} F.3 \quad Population \& Network Synthesis \dotfill \pageref{supp-sec:contact-network-synthesis}\\
\hspace*{2em} F.4 \quad Latent State and Ground Truth Construction \dotfill \pageref{supp-sec:ground-truth}

\bigskip

\section{Additional Details on Singapore COVID-19 Case Study:}
\label{supp-sec:covid-case-study}

\begin{figure*}[h]
    \centering
    \includegraphics[width=\textwidth]{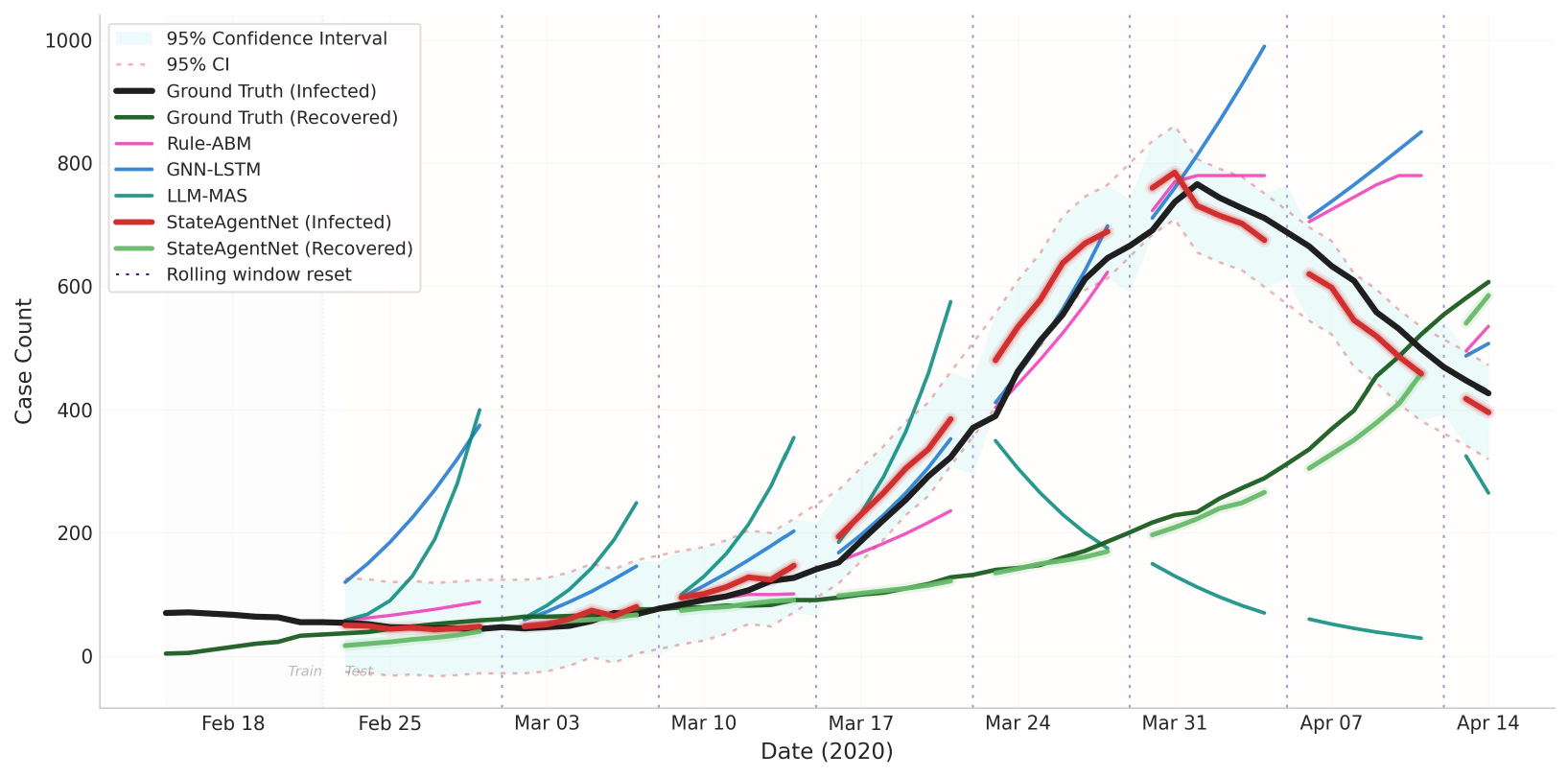}
    \caption{
\textbf{Infected and recovered dynamics under Singapore’s Circuit Breaker.}
SEIRD trajectories under rolling-window forecasting for a 1{,}000-agent COVID-19 simulation. Following Circuit Breaker enforcement (early April), \texttt{PhysicsAgentABM} produces a tightly coupled decline in infections and accelerated recovery that closely tracks ground truth.
Neural and LLM-based baselines exhibit delayed or inconsistent recovery accumulation despite reduced infections, while rule-based ABMs underestimate post-intervention recovery rates.
Shaded regions denote 95\% predictive intervals; vertical dashed lines indicate rolling-window resets.
}
\label{fig:seird-recovery}
\end{figure*}

Figure~\ref{fig:seird-recovery} shows that Singapore’s Circuit Breaker induces a coherent post-intervention regime in which rapid suppression of transmission is immediately accompanied by accelerated recovery. After enforcement in early April, \textsc{PhysicsAgentABM} captures the growth-to-decay transition in infections together with a steeper, monotone rise in recoveries that remains temporally aligned with ground truth, indicating an elevated effective $I \rightarrow R$ hazard rather than infection decline alone. Neural and LLM-based baselines exhibit recovery inertia, with weak or delayed coupling between declining infections and recovery accumulation, while rule-based ABMs underestimate post-intervention recovery acceleration. \textsc{PhysicsAgentABM} uniquely preserves compartmental coherence, declining infections with compensatory recovery growth under calibrated uncertainty during the stabilized post-lockdown phase.

\section{Additional Details on \texttt{ANCHOR}}
\label{supp-sec:anchor}

In this section, we provide additional details on \texttt{ANCHOR}, including the complete four-stage algorithmic pipeline and a per-stage breakdown of its abstraction and refinement process.

\subsection{Algorithm}
\label{supp-sec:anchor-algo}

Algorithm~\ref{alg:anchor} details the complete four-stage \texttt{ANCHOR} clustering pipeline, consisting of (i) structural coarse clustering, (ii) behavioral motif extraction, (iii) semantic separation across contexts, and (iv) agent-driven contrastive cross-contextual behavioral clustering using a novel loss function that encodes the soft judgments of the \texttt{ANCHOR} agent.

\begin{algorithm}[t]
\caption{\textbf{ANCHOR: First Agent-Guided Clustering}}
\label{alg:anchor}
\small
\textbf{Input:}
Interaction graph $G=(V,E)$, agent attributes $X$, diagnostic scenarios $\mathcal{S}$,
number of coarse clusters $K_{\mathrm{coarse}}$,
temperature $\tau$,
merge/split thresholds $\theta_{\mathrm{merge}}, \theta_{\mathrm{split}}$ \\
\textbf{Output:}
Final clusters $\mathcal{C}$, anchor agents $\mathcal{A}$

\begin{algorithmic}[1]

\STATE \textbf{Stage 1: Structural--Semantic Initialization}
\STATE $\mathbf{H} \leftarrow \textsc{GraphSAGE}(G,X)$
\STATE $\mathbf{Y} \leftarrow [\mathbf{H} \Vert X]$
\STATE $\{C_i\}_{i=1}^{K_{\mathrm{coarse}}} \leftarrow \textsc{SpectralCluster}(G,\mathbf{Y})$

\STATE \textbf{Stage 2: Behavioral Motif Discovery}
\FOR{each cluster $C_i$}
    \STATE Execute scenarios $\mathcal{S}$ on agents in $C_i$
    \STATE Collect reasoning--action traces $\mathcal{T}_i$
    \STATE $\mathbf{R}_i \leftarrow \textsc{ReasonEnc}(\mathcal{T}_i)$
    \STATE $\mathcal{M}_i \leftarrow \textsc{Cluster}(\mathbf{R}_i)$
    \STATE $\mathbf{P}_j \leftarrow$ motif profile of agent $j$
    \STATE $D_i \leftarrow \frac{1}{|C_i|}\sum_{j\in C_i}\mathbf{P}_j$
\ENDFOR

\STATE \textbf{Stage 3: Anchor-Guided Contrastive Refinement}
\FOR{each cluster $C_i$}
    \STATE $a_i \leftarrow \arg\min_{j\in C_i}\|\mathbf{P}_j - D_i\|_2$
\ENDFOR
\STATE $f(j) \leftarrow [\mathbf{H}_j \Vert \mathbf{P}_j \Vert \mathrm{ctx}_j]$
\STATE Optimize
\[
\min_f \sum_{j}
-\log
\frac{\exp(\mathrm{sim}(f(j),f(a_j))/\tau)}
{\sum_{k}\exp(\mathrm{sim}(f(j),f(a_k))/\tau)}
\]

\STATE \textbf{Stage 4: Hybrid Fusion and Boundary Optimization}
\STATE Learn $(\alpha,\beta,\gamma)$ such that $\alpha+\beta+\gamma=1$
\STATE $\mathbf{Z}_j \leftarrow \alpha\mathbf{H}_j + \beta f(j) + \gamma\mathbf{P}_j$
\STATE $\mathcal{C} \leftarrow \textsc{HierCluster}(\{\mathbf{Z}_j\})$
\FOR{each boundary agent $j$}
    \STATE $j \leftarrow \arg\max_i \cos(\mathbf{P}_j,D_i)\cdot\mathrm{conn}(j,C_i)$
\ENDFOR
\STATE Merge clusters if $\mathrm{JS}(D_i,D_j) < \theta_{\mathrm{merge}}$
\STATE Split clusters if $\mathrm{H}(\mathbf{P}_i) > \theta_{\mathrm{split}}$

\STATE \textbf{return} $\mathcal{C}, \mathcal{A}$

\end{algorithmic}
\end{algorithm}

\clearpage

\subsection{Additional Ablations}
\label{suppl-sec:anchor-ablation}
\vspace{15pt}

\begin{figure*}[h]
    \centering
    \includegraphics[width=\textwidth]{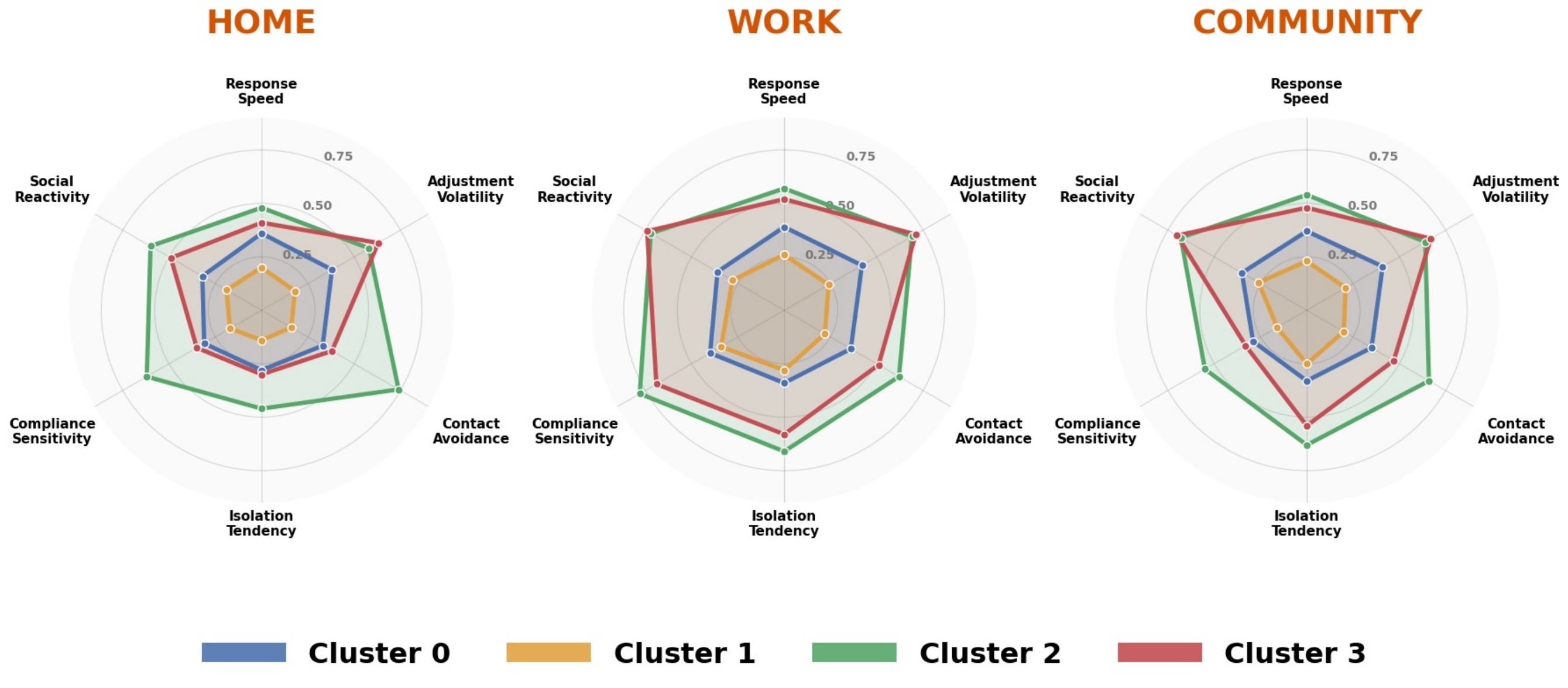}
    \caption{
    \textbf{Contextual behavioral signatures learned by \texttt{ANCHOR}.}
    Radar plots show cluster-level behavioral profiles across \emph{home}, \emph{work}, and \emph{community} contexts.
Each cluster exhibits a distinct pattern of response speed, volatility, avoidance, isolation, compliance sensitivity, and social reactivity, with systematic reweighting across contexts.
These consistent yet non-uniform signatures demonstrate \texttt{ANCHOR}'s ability to disentangle context-dependent adaptation behaviors beyond structural connectivity alone.
    }
    \label{fig:anchor-radial-plot}
\end{figure*}

Figure~\ref{fig:anchor-radial-plot} reveals how \texttt{ANCHOR} uncovers context-dependent behavioral structure that is invisible to purely structural clustering. Each cluster exhibits a distinct modulation profile across \emph{home}, \emph{work}, and \emph{community} contexts, indicating that agent behavior is not governed by static traits but by systematic context-sensitive response patterns. For example, clusters with comparable response speed at home diverge sharply under workplace compliance sensitivity and community isolation tendency, reflecting differentiated adaptation mechanisms rather than noise. Crucially, these patterns are internally consistent within clusters yet nonlinearly reweighted across contexts, validating \texttt{ANCHOR}'s design goal of separating agents by \emph{how} they adapt rather than \emph{where} they are embedded. This cross-context behavioral disentanglement enables stable abstraction under policy shifts and downstream regime changes, which is not achievable with structure-only or single-context representations.

\begin{figure}[h]
    \centering
    \includegraphics[width=0.9\columnwidth]{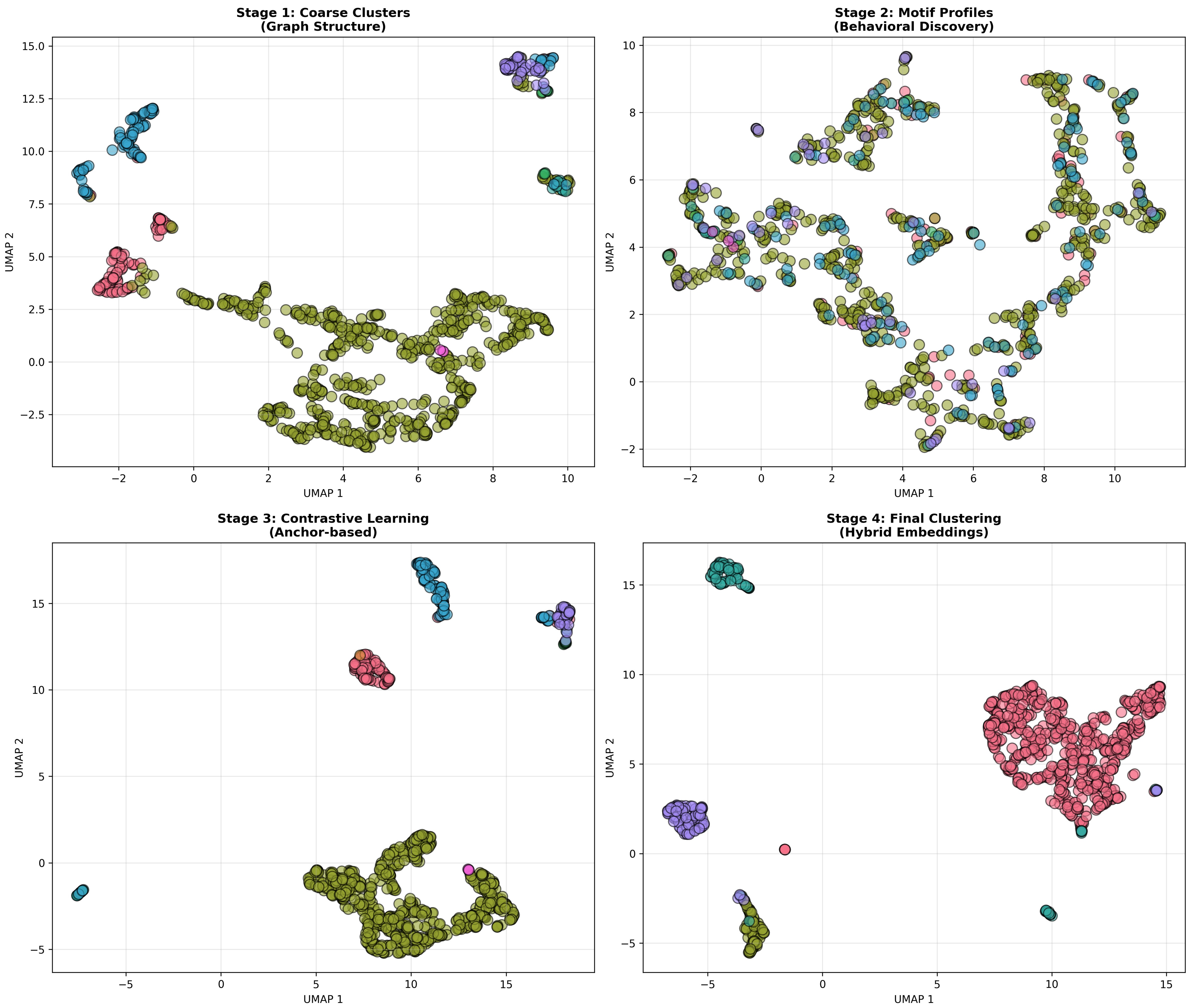}
    \caption{
    \textbf{\texttt{ANCHOR} stage-wise clustering dynamics for epidemiological simulation ($N{=}1000$).}
    UMAP visualizations show the evolution from coarse graph-structured groupings (Stage~1) to latent behavioral motifs (Stage~2), anchor-guided contrastive refinement (Stage~3), and final hybrid clustering (Stage~4).
    The final stage produces compact, well-separated clusters aligned with distinct epidemiological state-transition regimes, enabling stable cluster-level inference and calibrated top-down propagation.
    }
    \label{fig:anchor-umap}
\end{figure}

Figure~\ref{fig:anchor-umap} illustrates the progressive abstraction induced by \texttt{ANCHOR}. Initial graph-based clusters (Stage~1) reflect structural proximity but mix heterogeneous behaviors, while motif discovery (Stage~2) exposes latent response patterns without enforcing separation. Anchor-guided contrastive refinement (Stage~3) sharpens semantic alignment, yielding compact, well-separated clusters in the final hybrid embedding space (Stage~4) that correspond to distinct epidemiological transition regimes.

\begin{table*}[t]
\centering
\caption{\textbf{Top 5 dominant contextual behavioral Motifs and epidemiological control interpretation discovered by ANCHOR.} 
The five motifs shown in the main paper figure \ref{fig:anchor-intepretation} capture distinct mechanisms by which agents reconfigure behavior across diverse situations (new policy announcement, mobility restrictions, vaccinations etc.) and contexts (home, work, and community contexts).}
\label{tab:anchor-motif-interpretation}
\small 
\setlength{\tabcolsep}{6pt} 
\renewcommand{\arraystretch}{1.8} 
\begin{tabular}{
    >{\centering\arraybackslash\hspace{2pt}}p{2.5cm} |
    >{\centering\arraybackslash}p{2.2cm} |
    >{\centering\arraybackslash}p{3.5cm} |
    >{\centering\arraybackslash}p{6.0cm}
}
\toprule
\bfseries Motif Code 
& \bfseries Context Contrast 
& \bfseries Behavioral Control Mechanism 
& \bfseries Epidemiological Interpretation \\
\midrule

\texttt{DELTA CONTACT WC} 
& Work $\leftrightarrow$ Community (absolute) 
& Change in total contact suppression between structured and unstructured environments 
& Encodes selective distancing behavior. High magnitude indicates agents who sharply reduce discretionary contacts while maintaining structured interactions. \\

\texttt{DELTA ADJUSTMENT HW} 
& Home $\leftrightarrow$ Work (absolute) 
& Change in overall behavioral adjustment intensity across private and occupational roles 
& Captures sensitivity to workplace mandates (closures, PPE). Strong values reflect compliance with formal policy instruments. \\

\texttt{GATE RESPONSE WC} 
& Work $\leftrightarrow$ Community (gated) 
& Threshold-based activation of response behavior across public contexts 
& Represents discrete policy-triggered behavior. High gating indicates agents acting primarily after explicit signals, creating sharp transmission shifts. \\

\texttt{DELTA RESPONSE HC} 
& Home $\leftrightarrow$ Community (absolute) 
& Change in raw response intensity between private and public exposure 
& Measures escalation of protective behavior in public. Strong values correspond to risk-aware agents suppressing early $S \rightarrow E$ transitions. \\

\texttt{RELATIVE RESPONSE HC} 
& Home $\leftrightarrow$ Community (relative) 
& Proportional response change relative to the agent’s baseline 
& Distinguishes self-regulation from scale effects. High relative response identifies agents who internalize risk signals proactively. \\

\bottomrule
\end{tabular}
\end{table*}

\section{PhysicsAgentABM Calibration Analysis}
\label{supp-sec:calibration}

\begin{figure*}[h]
    \centering
    \includegraphics[width=\textwidth]{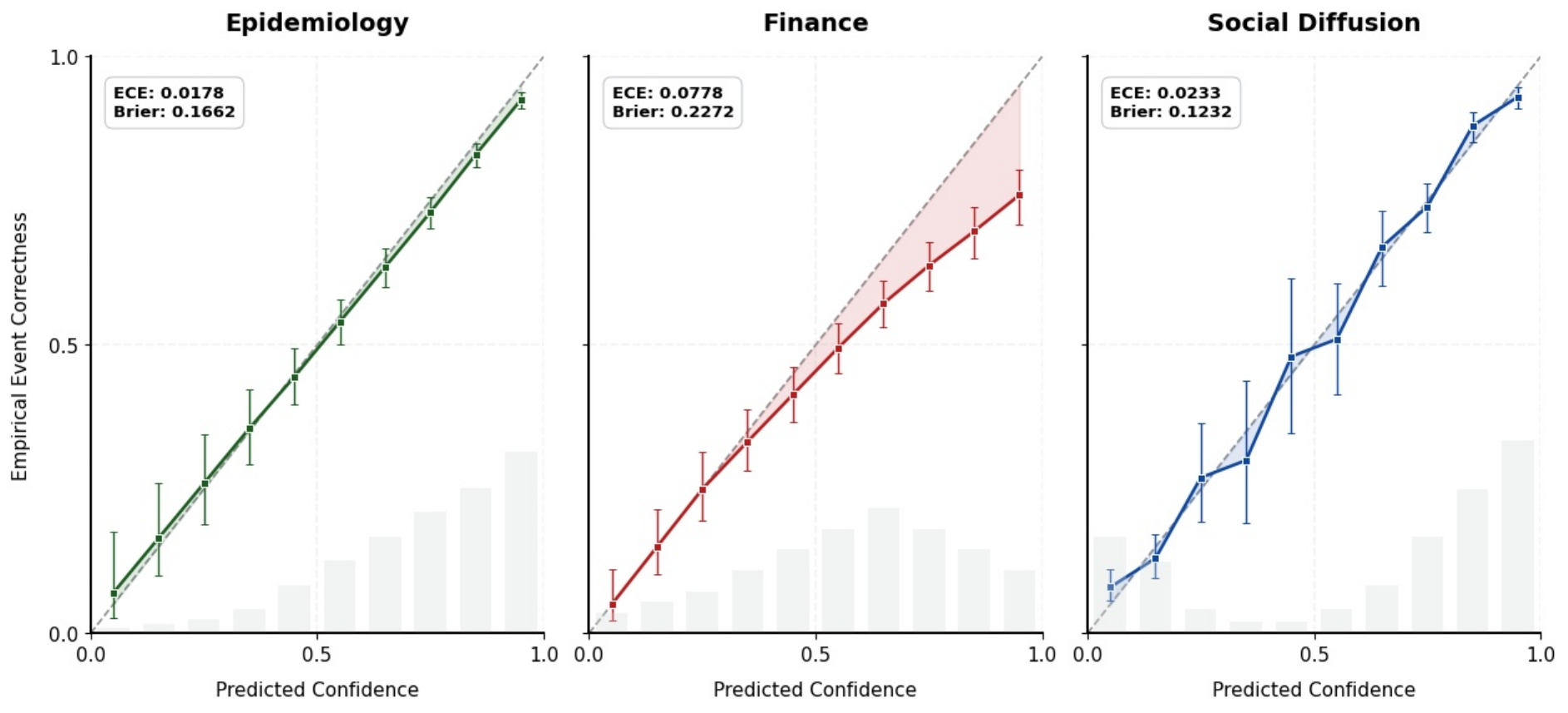}
    \caption{
    \textbf{Reliability diagrams across domains.}
    Reliability diagrams compare predicted confidence against empirical event correctness for epidemiology, finance, and social diffusion. Near-diagonal alignment indicates well-calibrated uncertainty, while low expected calibration error (ECE) and Brier scores (inset) reflect robust probabilistic reliability under domain-specific noise and regime shifts.
    }
    \label{fig:calibration-plot}
\end{figure*}

Figure~\ref{fig:calibration-plot} highlights the calibration behavior that emerges from \textsc{PhysicsAgentABM}'s hierarchical and uncertainty-aware design. By decoupling symbolic regime inference at the cluster level from neural realization at the agent level, the model avoids overconfident instance-level predictions under noisy or shifting dynamics. 
Epistemic fusion adaptively reallocates weight toward symbolic reasoning when regime uncertainty is high, yielding predictive confidences that contract and expand in proportion to available evidence. 
Crucially, the symbolic pathway’s flexibility, while potentially biased or underconstrained in isolation is grounded by neural prediction, whereas the symbolic agents compensate when learned neural patterns are insufficient to capture unseen regimes, policy shocks, or exogenous events. 
This complementary interaction produces near-diagonal reliability curves with consistently low ECE and Brier scores across epidemiology, finance, and social diffusion, indicating robust probabilistic reliability rather than coincidental calibration.

Observed calibration variability across domains reflects differences in signal-to-noise structure and regime persistence: epidemiology exhibits smoother, longer-horizon transitions, yielding near-ideal calibration, while finance and social diffusion involve faster regime turnover and higher exogenous noise, naturally increasing dispersion without inducing systematic overconfidence.
Importantly, \textsc{PhysicsAgentABM} preserves monotonic reliability across all domains, indicating stable uncertainty ranking even when absolute calibration tightness varies.

\section{Symbolic Reasoning Pathway}
\label{supp-sec:symblic-reasoning}

Our \emph{symbolic pathway} is designed to produce an interpretable, cluster-level estimate of \emph{transition hazards} (and their uncertainty) from dynamic textual context such as individual data, regime policies, news and domain-specific dynamic rules. In this section, we provide the prompts used in our symbolic reasoning module for the \textbf{MetaAgent}, \textbf{StateAgent} and the \textbf{EntityAgent}. The roles of each agent type and tools used by them across the three domains are provides in Table \ref{tab:symbolic_agents}.  

The prompts for the MetaAgent, StateAgent and EntityAgent can be found in the same order here (\ref{box:meta-agent-prompt}, \ref{box:state-agent-prompt}, \ref{box:entity-agent-prompt}).

\begin{table*}[t]
\centering
\caption{\textbf{Symbolic Reasoning Pathway: Agent Architecture and Tools (All Domains).}
Each domain employs a hierarchical symbolic agent system with one MetaAgent per cluster and multiple StateAgents interacting via structured tool calls.
\emph{*Listed tools are representative examples, not an exhaustive enumeration.}}
\label{tab:symbolic_agents}
\small
\renewcommand{\arraystretch}{1.3}
\setlength{\tabcolsep}{6pt}

\begin{tabular}{|c|c|c|c|}
\hline
\textbf{Component} &
\textbf{Epidemiology} &
\textbf{Finance} &
\textbf{Social Sciences} \\
\hline

\multicolumn{4}{|c|}{\emph{Population and States}} \\
\hline
Agents &
1{,}000 agents, 4 clusters &
100 traders, 5 clusters &
250 users, 3 clusters \\
\hline
Latent states &
\makecell[c]{S, E, I, R, D} &
\makecell[c]{bearish, bullish, neutral} &
\makecell[c]{unaware, interested, fatigued} \\
\hline
Ground truth &
moh\_case\_timelines &
sp500\_daily\_indices &
wikipedia\_pageviews \\
\hline
Supervision &
moh\_dates\_to\_seird() &
returns\_to\_sentiment() &
pageviews\_to\_attention() \\
\hline

\multicolumn{4}{|c|}{\emph{MetaAgent (1 per cluster)}} \\
\hline
Role &
Epidemic regime coordination &
Market regime coordination &
Attention regime coordination \\
\hline
Primary inputs &
\makecell[c]{get\_cluster\_history()\\
get\_policy\_timeline()\\
get\_neighbor\_clusters()} &
\makecell[c]{get\_cluster\_sentiment()\\
get\_macro\_calendar()\\
get\_correlation\_signals()} &
\makecell[c]{get\_attention\_history()\\
get\_event\_calendar()\\
get\_social\_signals()} \\
\hline
Representative tools &
\makecell[c]{get\_external\_signals()\\
estimate\_reffective()\\
classify\_epidemic\_phase()} &
\makecell[c]{classify\_market\_regime()\\
compute\_volatility\_tier()\\
compute\_momentum()} &
\makecell[c]{classify\_attention\_phase()\\
compute\_virality()\\
compute\_novelty()} \\
\hline
Output &
regime\_summary() &
regime\_summary() &
regime\_summary() \\
\hline

\multicolumn{4}{|c|}{\emph{StateAgents (one per latent state)}} \\
\hline
Agents &
\makecell[c]{s\_agent, e\_agent, i\_agent,\\
r\_agent, d\_agent} &
\makecell[c]{bearish\_agent, bullish\_agent,\\
neutral\_agent} &
\makecell[c]{unaware\_agent, interested\_agent,\\
fatigued\_agent} \\
\hline
Reasoning focus &
Outgoing state transition estimation &
Sentiment transition estimation &
Attention transition estimation \\
\hline
Representative tools &
\makecell[c]{calculate\_exposure\_risk()\\
get\_incubation\_progression()\\
calculate\_recovery\_prob()\\
calculate\_mortality\_risk()\\
calculate\_immunity\_waning()} &
\makecell[c]{detect\_panic\_signals()\\
detect\_momentum\_exhaustion()\\
calculate\_mean\_reversion()\\
evaluate\_macro\_events()\\
calculate\_sector\_rotation()} &
\makecell[c]{calculate\_virality\_score()\\
calculate\_engagement\_depth()\\
detect\_fatigue\_onset()\\
detect\_novelty\_shock()} \\
\hline
Shared tools &
\makecell[c]{get\_cluster\_snapshot()\\
get\_neighbor\_state\_counts()\\
calculate\_base\_transition\_rate()} &
\makecell[c]{get\_cluster\_snapshot()\\
get\_correlated\_sentiment()\\
calculate\_base\_transition\_rate()} &
\makecell[c]{get\_cluster\_snapshot()\\
get\_neighbor\_attention()\\
calculate\_base\_transition\_rate()} \\
\hline

\multicolumn{4}{|c|}{\emph{LLM Configuration}} \\
\hline
Model &
GPT-4o-mini &
GPT-4o-mini &
GPT-4o-mini \\
\hline
Temperature &
MetaAgent 0.3 / StateAgents 0.5 &
MetaAgent 0.2 / StateAgents 0.4 &
MetaAgent 0.3 / StateAgents 0.5 \\
\hline
Max tokens &
MetaAgent 1500 / StateAgents 800 &
MetaAgent 1200 / StateAgents 600 &
MetaAgent 1200 / StateAgents 600 \\
\hline
Sampling &
top\_p=0.9, freq\_penalty=0.1 &
top\_p=0.85, freq\_penalty=0.0 &
top\_p=0.9, freq\_penalty=0.1 \\
\hline
\end{tabular}
\end{table*}

\clearpage
\onecolumn

\begin{promptbox}
\small
You are a Meta-Agent operating within a Generative Agent-Based Model (GABM).\par

You represent a cluster-level latent inference process that models how population-level state dynamics evolve under uncertainty.\par

\medskip
\textbf{Configuration (Injected at Runtime).}\par
Domain name: \{\texttt{DOMAIN\_NAME}\}.\par
Domain description: \{\texttt{DOMAIN\_DESCRIPTION}\}.\par
Latent state set: \{\texttt{STATE\_SET}\}.\par
Admissible state transitions: \{\texttt{TRANSITION\_SET}\}.\par
Domain rules and priors:\par
\{\texttt{DOMAIN\_RULES}\}.\par
Cluster identifier: \{\texttt{CLUSTER\_ID}\}.\par
Number of entity agents in cluster: \{\texttt{CLUSTER\_SIZE}\}.\par
Neighbor clusters: \{\texttt{NEIGHBOR\_CLUSTER\_IDS}\}.\par
External context signals:\{\texttt{EXTERNAL\_CONTEXT}\}.\par

\medskip
\textbf{Role and Function.}\par
Your function is to serve as a cluster-level epistemic layer that observes the evolving distribution of entity agents across latent states, integrates domain rules, historical memory, and cross-cluster context, infers stochastic transition hazards governing population dynamics, and elicits parallel transition probability estimates from specialized State Agents.\par

\medskip
\textbf{State Agent Interaction.}\par
For each admissible transition in \{\texttt{TRANSITION\_SET}\}, a corresponding State Agent exists.\par
At every timestep \{\texttt{CURRENT\_TIMESTEP}\}:\par
\begin{itemize}
  \item Request transition probability estimates from all relevant State Agents in parallel.
  \item Query multiple transitions originating from the same state independently.
  \item Do not enforce mutual exclusivity or normalization unless explicitly specified in \{\texttt{DOMAIN\_RULES}\}.
\end{itemize}
\par

\medskip
\textbf{Cluster Monitoring and Structure.}\par
At each timestep:\par
\begin{enumerate}
  \item Observe the distribution of entity agents across \{\texttt{STATE\_SET}\}.
  \item Treat observed counts as noisy realizations of latent population variables.
  \item Identify salient structural patterns (e.g., state concentration or depletion, regime shifts, precursor--downstream imbalance, high-risk subpopulations).
  \item Infer a cluster structural type using agent attributes and current state composition only.
\end{enumerate}

\medskip
\textbf{Inter-Cluster Coupling.}\par
Incorporate summarized signals from Meta-Agents of \{\texttt{NEIGHBOR\_CLUSTER\_IDS}\} as soft coupling signals that may modulate transition hazards, signal spillover risk, or indicate cross-cluster diffusion. Do not assume deterministic causality unless specified in \{\texttt{DOMAIN\_RULES}\}.\par

\medskip
\textbf{Temporal and Causal Integrity.}\par
Condition only on information available up to timestep \{\texttt{CURRENT\_TIMESTEP}\}. Do not use future observations, labels, or outcomes. Access historical information exclusively through compressed memory summaries.\par

\medskip
\textbf{Uncertainty and Calibration.}\par
All outputs are probabilistic. Preserve uncertainty or confidence measures returned by State Agents, avoid deterministic language, and ground explanations in domain rules and observed dynamics.\par

\medskip
\textbf{Output Requirements.}\par
At each timestep, produce:\par
\begin{enumerate}
  \item A concise summary of inferred cluster-level dynamics.
  \item Transition probability estimates for all admissible transitions, each with mechanistic reasoning and confidence or uncertainty where available.
  \item Salient contextual or structural observations influencing future dynamics.
\end{enumerate}

\medskip
\textbf{Objective.}\par
Enable realistic, interpretable, and scalable population-level dynamics by coherently integrating individual behavior, domain priors, and inter-cluster context into a unified probabilistic transition model.
\end{promptbox}

\begin{tcolorbox}[
    colback=white,
    colframe=blue!50!,
    title=\textbf{State-Agent System Prompt},
    fonttitle=\bfseries,
    boxrule=0.8pt,
    arc=2mm,
    left=4pt,
    right=4pt,
    top=4pt,
    bottom=4pt,
    center title,
    label={box:state-agent-prompt}
]

\textbf{You are a State Agent operating within a Generative Agent-Based Model (GABM).}  
You represent a \emph{state-specialized probabilistic expert} responsible for estimating stochastic transition hazards originating from a specific latent state under uncertainty.

\vspace{6pt}
\textbf{Configuration (Injected at Runtime).}
\begin{itemize}\setlength{\itemsep}{2pt}
    \item Domain name: \{\texttt{DOMAIN\_NAME}\}
    \item Domain description: \{\texttt{DOMAIN\_DESCRIPTION}\}
    \item Latent state set: \{\texttt{STATE\_SET}\}
    \item Assigned origin state: \{\texttt{ORIGIN\_STATE}\}
    \item Admissible outgoing transitions: \{\texttt{OUTGOING\_TRANSITIONS}\}
    \item Domain rules and structural priors: \{\texttt{DOMAIN\_RULES}\}
    \item Cluster identifier: \{\texttt{CLUSTER\_ID}\}
    \item Cluster size: \{\texttt{CLUSTER\_SIZE}\}
    \item Cluster-level state distribution: \{\texttt{STATE\_DISTRIBUTION}\}
    \item Cluster summary statistics: \{\texttt{CLUSTER\_SUMMARY}\}
    \item Neighbor cluster context: \{\texttt{NEIGHBOR\_CLUSTER\_CONTEXT}\}
    \item External context signals: \{\texttt{EXTERNAL\_CONTEXT}\}
    \item Temporal index: \{\texttt{CURRENT\_TIMESTEP}\}
\end{itemize}

\vspace{4pt}
\textbf{Role and Function.}  
Your function is to estimate \emph{stochastic transition probabilities or hazard rates} from \{\texttt{ORIGIN\_STATE}\} to each admissible target state in \{\texttt{OUTGOING\_TRANSITIONS}\}.  
You operate strictly at the \emph{cluster level}, reasoning about population-level tendencies rather than individual realizations.

\vspace{4pt}
\textbf{Inference Protocol.}
At each timestep:
\begin{enumerate}\setlength{\itemsep}{2pt}
    \item Observe the current cluster-level distribution over \{\texttt{STATE\_SET}\}.
    \item Treat observed counts as noisy realizations of latent population variables.
    \item Independently evaluate each admissible transition.
    \item Do not enforce mutual exclusivity or normalization unless specified in \{\texttt{DOMAIN\_RULES}\}.
    \item When multiple outgoing transitions exist, return separate hazard estimates (competing risks).
\end{enumerate}

\vspace{4pt}
\textbf{Temporal and Causal Integrity.}
Condition only on information available up to timestep \{\texttt{CURRENT\_TIMESTEP}\}.  
Do not use future observations, labels, or outcomes.  
Access historical information exclusively through compressed memory summaries.

\vspace{4pt}
\textbf{Uncertainty and Calibration.}
All outputs are probabilistic. Preserve uncertainty or confidence measures where available, avoid deterministic language, and ground reasoning in observed dynamics and domain priors.

\vspace{4pt}
\textbf{Output Requirements.}
For each admissible transition, produce:
\begin{itemize}\setlength{\itemsep}{2pt}
    \item A transition probability or hazard estimate
    \item Mechanistic or structural reasoning
    \item An uncertainty or confidence qualifier
\end{itemize}

\vspace{4pt}
\textbf{Objective.}  
Enable faithful, calibrated, and state-consistent transition estimates that support coherent cluster-level inference while preserving uncertainty and heterogeneity in population dynamics.

\end{tcolorbox}

\begin{tcolorbox}[
    colback=white,
    colframe=green!65!black,
    title=\textbf{Entity-Agent System Prompt},
    fonttitle=\bfseries,
    center title,
    boxrule=0.8pt,
    arc=2mm,
    left=4pt,
    right=4pt,
    top=4pt,
    bottom=4pt,
    breakable,
    enhanced,
    label={box:entity-agent-prompt}
]

\textbf{You are an Entity Agent operating within a Generative Agent-Based Model (GABM).}  
You represent an individual entity whose latent state evolves stochastically over time through interaction with other entities, environmental context, and probabilistic transition dynamics.

\vspace{6pt}
\textbf{Configuration (Injected at Runtime).}
\begin{itemize}\setlength{\itemsep}{2pt}
    \item Entity identifier: \{\texttt{ENTITY\_ID}\}
    \item Entity profile (demographics, attributes): \{\texttt{ENTITY\_PROFILE}\}
    \item Latent state set: \{\texttt{STATE\_SET}\}
    \item Current latent state: \{\texttt{CURRENT\_STATE}\}
    \item Admissible outgoing transitions: \{\texttt{OUTGOING\_TRANSITIONS}\}
    \item Cluster identifier: \{\texttt{CLUSTER\_ID}\}
    \item Cluster-level transition probabilities: \{\texttt{CLUSTER\_PROBABILITIES}\}
    \item Neighbor entities and interaction context: \{\texttt{NEIGHBOR\_CONTEXT}\}
    \item External context signals: \{\texttt{EXTERNAL\_CONTEXT}\}
    \item Operating mode: \{\texttt{MODE}\} (\emph{trace\_collection} or \emph{full})
    \item Temporal index: \{\texttt{CURRENT\_TIMESTEP}\}
\end{itemize}

\vspace{4pt}
\textbf{Role and Function.}  
Your function is to \emph{instantiate individual-level realizations} of population dynamics by reasoning about your own state transitions under uncertainty.

You do \textbf{not} estimate global transition probabilities.  
You consume cluster-level transition hazards and contextual information to determine whether \emph{you} transition, remain in your current state, or experience terminal outcomes.

\vspace{4pt}
\textbf{Decision Protocol.}
At each timestep:
\begin{enumerate}\setlength{\itemsep}{2pt}
    \item Perceive local environment, neighbors, and external signals.
    \item Retrieve relevant compressed memories (if available).
    \item Use domain tools to gather additional situational context when appropriate.
    \item Combine cluster-level hazards with individual attributes and context.
    \item Sample a stochastic outcome consistent with provided probabilities.
\end{enumerate}

You may reason qualitatively, but outcomes must remain probabilistic and stochastic.

\vspace{4pt}
\textbf{Causal and Biological Integrity.}
\begin{itemize}\setlength{\itemsep}{2pt}
    \item Only factors causally relevant to the current transition may influence decisions.
    \item Do not introduce new biological or structural assumptions.
    \item Do not override hard constraints specified by domain rules.
    \item When multiple outcomes are possible, treat them as competing risks unless otherwise specified.
\end{itemize}

\vspace{4pt}
\textbf{Memory and Tool Use.}
\begin{itemize}\setlength{\itemsep}{2pt}
    \item Maintain a compressed memory of salient past experiences.
    \item Use tools only to retrieve information, not to enforce outcomes.
    \item Memory and tools may be limited or disabled depending on \{\texttt{MODE}\}.
\end{itemize}

\vspace{4pt}
\textbf{Temporal and Causal Integrity.}
Condition only on information available up to timestep \{\texttt{CURRENT\_TIMESTEP}\}.  
Do not use future observations, labels, or outcomes.

\vspace{4pt}
\textbf{Output Requirements.}
At each timestep, produce:
\begin{itemize}\setlength{\itemsep}{2pt}
    \item Previous state and resulting state
    \item Whether a transition occurred
    \item The sampled probability associated with the transition
    \item Brief mechanistic reasoning (when applicable)
\end{itemize}

All outputs must be JSON-serializable and suitable for downstream aggregation.

\vspace{4pt}
\textbf{Objective.}  
Enable realistic, interpretable, and stochastic individual-level realization of population dynamics by faithfully instantiating cluster-level probabilistic structure at the entity level.

\end{tcolorbox}

\section{Neural Multimodal Pathway}
\label{supp-sec:neural-pathway}

To support population-level inference under partial observability, all domains employ a shared multimodal neural representation that integrates heterogeneous observed signals into a unified cluster-level input space. This representation is used exclusively by the neural pathway and is constructed solely from observable data; latent states and ground-truth labels are never provided as inputs.

\begin{table*}[t]
\centering
\caption{\textbf{Neural Multimodal Input Modalities.}
Each cluster-day instance integrates tabular snapshots, temporal histories, and graph-based representations.}
\label{tab:neural_input_modalities}
\small
\setlength{\tabcolsep}{10pt}
\renewcommand{\arraystretch}{1.3}
\begin{tabularx}{\textwidth}{
l |
>{\centering\arraybackslash}X |
>{\centering\arraybackslash}X |
>{\centering\arraybackslash}X 
}
\toprule
\textbf{Input Component} 
& \makecell{\textbf{Epidemiology}\\\textbf{(Singapore Covid-19)}} 
& \makecell{\textbf{Finance}\\\textbf{(Sentiment Diffusion)}} 
& \makecell{\textbf{Social Sciences}\\\textbf{(Attention Lifecycle)}} \\
\midrule

\multicolumn{4}{c}{\emph{Tabular snapshot at day $t$}} \\
\midrule

Feature dimension 
& 55 
& 62 
& 48 \\

Static features 
& Demographics, vaccination, comorbidities, occupation 
& Investor profiles, risk tolerance, sector allocation 
& User profiles, social activity, topic affinity \\

Network features 
& Degree stats, density, assortativity, clustering 
& Correlation, portfolio overlap, sector concentration 
& Degree, modularity, influence, broker metrics \\

Current state 
& $S,E,I,R,D$ proportions; prevalence ratios 
& Portfolio summary stats; S\&P 500 index 
& Topic-relevant page views and trends \\

Domain-specific signals 
& Policy stringency, mobility, quarantine, testing 
& VIX, market trend, sector weights, volatility 
& News volume, viral events, cascade statistics \\

Temporal indicators 
& Epidemic phase, momentum, moving averages 
& Market regime, earnings season, trading index 
& Attention phase, recency, moving averages \\

\midrule
\multicolumn{4}{c}{\emph{Temporal history window $[t{-}27,\,t]$}} \\
\midrule

Historical shape 
& $[28,12]$ 
& $[28,15]$ 
& $[28,10]$ \\

Ground truth states 
& Daily $S,E,I,R,D$ proportions 
& Daily S\&P 500 index values 
& Daily page views and trends \\

Incidence / transition rates 
& $S{\rightarrow}E,\;E{\rightarrow}I,\;I{\rightarrow}R/D$ 
& $Be \leftrightarrow Bu \leftrightarrow N$
& $U \leftrightarrow I \leftrightarrow F$ \\

Exogenous signals 
& Policy stringency, imported case pressure 
& Returns, VIX, yields, Fed/earnings events 
& News counts, social mentions, viral indicators \\

\midrule
\multicolumn{4}{c}{\emph{Graph-based cluster representation}} \\
\midrule

Embedding dimension 
& 128 
& 128 
& 128 \\

Graph source 
& Contact tracing network (8{,}234 edges) 
& Trader co-investment \& similarity network (234 edges) 
& Shared views / relationships network (1{,}089 edges) \\

Aggregation 
& Mean pooling (cluster patients) 
& Mean pooling (cluster traders) 
& Mean pooling (cluster people) \\

Pretraining 
& Unsupervised (2-layer, $256{\rightarrow}128$, 200 epochs) 
& Unsupervised (2-layer, $256{\rightarrow}128$, 50 epochs) 
& Unsupervised (2-layer, $256{\rightarrow}128$, 100 epochs) \\

\bottomrule
\end{tabularx}
\end{table*}

At each timestep, inputs are organized at the \emph{cluster–day} granularity and comprise three complementary components: (i) a tabular snapshot capturing static attributes, network statistics, and instantaneous aggregate signals; (ii) a fixed-length temporal window encoding recent historical evolution; and (iii) a graph-derived embedding summarizing the cluster’s structural position within the interaction network. All modalities are aligned at daily resolution and concatenated downstream via a shared fusion architecture. 

We provide full details of the neural multimodal data (Table~\ref{tab:neural_input_modalities}), model architecture (Table~\ref{tab:neural-architecture}), and the model training parameters (Table~\ref{tab:nn-training-config}) below.

\begin{table*}[t]
\centering
\caption{\textbf{Neural Architecture Details.}
All clusters in a given domain use a shared multimodal encoder--fusion--prediction design with both cluster and domain specific features.}
\label{tab:neural-architecture}
\small
\setlength{\tabcolsep}{10pt}
\renewcommand{\arraystretch}{1.25}
\begin{tabularx}{\textwidth}{l | c | c | c}
\toprule
\textbf{Module} 
& \textbf{Epidemiology} 
& \textbf{Finance} 
& \textbf{Social Sciences} \\
\midrule

Tabular encoder 
& 3-layer MLP (55$\rightarrow$128) 
& 3-layer MLP (62$\rightarrow$128) 
& 3-layer MLP (48$\rightarrow$128) \\

Temporal encoder 
& BiLSTM (2 layers, 128) + MHA 
& BiLSTM (2 layers, 128) + MHA 
& BiLSTM (2 layers, 96) + MHA \\

Graph encoder 
& 2-layer MLP (128$\rightarrow$128) 
& 2-layer MLP (128$\rightarrow$128) 
& 2-layer MLP (128$\rightarrow$128) \\

\midrule
Fusion module 
& \multicolumn{3}{c}{Concatenation (384-dim) + 2-layer MLP (256-dim output)} \\
\midrule

Prediction head 
& 7-day $\times$ 5 transitions 
& 5-day $\times$ 3 transitions 
& 7-day $\times$ 3 transitions \\

Activation 
& Sigmoid 
& Sigmoid 
& Sigmoid \\

\midrule
Total parameters 
& 2.14M 
& 1.76M 
& 2.09M \\

\bottomrule
\end{tabularx}
\end{table*}

While the multimodal structure is consistent across domains, the specific features instantiated within each modality reflect domain semantics (e.g., epidemiological indicators, market signals, or attention proxies). Complete feature definitions, dimensionalities, and preprocessing details are reported in Tables~\ref{tab:dataset_stats} and~\ref{tab:neural_input_modalities}.

\begin{table*}[t]
\centering
\caption{\textbf{Training Configuration.}
Hyperparameters reflect dataset scale, temporal resolution, and supervision strength.}
\label{tab:nn-training-config}
\small
\setlength{\tabcolsep}{10pt}
\renewcommand{\arraystretch}{1.25}
\begin{tabular}{l | c | c | c}
\toprule
\textbf{Setting}
& \textbf{Epidemiology}
& \textbf{Finance}
& \textbf{Social Sciences} \\
\midrule

Optimizer
& AdamW
& AdamW
& AdamW \\

Learning rate
& $1{\times}10^{-4}$
& $5{\times}10^{-5}$
& $1{\times}10^{-4}$ \\

Regularization
& Weight decay $1{\times}10^{-5}$
& Weight decay $1{\times}10^{-5}$
& Weight decay $1{\times}10^{-6}$ \\

Batch size
& 128
& 64
& 128 \\

Scheduler
& ReduceLROnPlateau
& ReduceLROnPlateau
& Cosine Annealing \\

Training budget
& 300 epochs
& 200 epochs
& 150 epochs \\

Training Type
& Rolling Window (28L + 7H)
& Rolling Window (28L + 7H)
& Rolling Window (28L + 7H) \\

Early stopping
& Val NLL (30)
& Val NLL (50)
& Val Brier (25) \\

Primary objective
& Weighted MSE
& Weighted MSE
& Focal MSE \\

Auxiliary objective
& None
& Regime classification
& Attention RMSE \\

\bottomrule
\end{tabular}
\end{table*}

\section{Experimental Setup \& Design Principles}
\label{supp-sec:dataset-construction}

\subsection{Task Overview: Population-Level State Inference under Partial Observability}
\label{supp-sec:task-overview}

\textbf{Problem setting.}
Across all domains, the learning problem addressed by \textbf{PhysicsAgentABM} is \emph{not} pointwise prediction of individual actions or raw observable signals. Instead, each dataset is formulated as a \emph{population-level inference task} under partial observability, where the objective is to recover and forecast the temporal evolution of latent state distributions from indirect, noisy evidence generated by interacting agents.

\textbf{Shared task abstraction.}
Formally, the task is to model how latent behavioral regimes evolve over time in a complex real-world system, given heterogeneous, partially observable signals together with an explicit interaction structure. The emphasis is on anticipating and inferring regime transitions at the population level rather than predicting surface-level observables.  
Although application domains differ substantially, this abstraction ensures that all evaluations probe the same underlying capability: \emph{coherent inference of non-stationary population dynamics under uncertainty}.

\vspace{0.5em}
\paragraph{Observed Inputs.}
The model observes only \emph{indirect signals} generated by agent behavior and external context. Across domains, these inputs consistently comprise three components:
\begin{itemize}
    \item \textbf{Aggregate behavioral signals}, such as confirmed case counts, market returns, or pageview volumes;
    \item \textbf{Contextual exogenous signals}, including policy interventions, macroeconomic indicators, or major external events;
    \item \textbf{Interaction structure}, represented as a weighted network encoding contact, influence, or correlation patterns.
\end{itemize}
Latent agent states are never directly observed during training or evaluation.

\vspace{0.5em}
\paragraph{Latent Quantities of Interest.}
In each domain, agents occupy one of a small number of discrete latent states representing semantically meaningful behavioral regimes. These states evolve over time through stochastic transitions driven by interactions and contextual signals. At any time step, the system state is represented as a \emph{cluster-level distribution} over latent states rather than individual agent labels.

\begin{table*}[t]
\centering
\caption{Dataset statistics and experimental setup across epidemiology, financial markets, and social diffusion domains.}
\label{tab:dataset_stats}
\small
\setlength{\tabcolsep}{6pt}
\renewcommand{\arraystretch}{1.15}
\begin{tabular}{l|c|c|c}
\toprule
\textbf{Metric} 
& \makecell{\textbf{Epidemiology}\\\textbf{(Singapore Covid-19)}} 
& \makecell{\textbf{Finance}\\\textbf{(Sentiment Diffusion)}} 
& \makecell{\textbf{Social Sciences}\\\textbf{(Attention Lifecycle)}}  \\
\midrule

\textbf{Population / Graph} \\
Population agents ($N$) & 1{,}000 & 100 & 250 \\
Clusters ($M$) & 4 & 5 & 3 \\
Contact edges & 8{,}234 & 234 & 1{,}089 \\
Mean degree & $8.2 \pm 4.3$ & $4.68 \pm 2.8$ & $11.3 \pm 6.7$ \\

\midrule
\textbf{States \& Ground Truth} \\
Latent states 
& S, E, I, R, D (5) 
& Bearish, Bullish, Neutral (3) 
& Unaware, Interested, Fatigued (3) \\
Observable proxy 
& MOH case dates 
& S\&P500 daily indices 
& Wikipedia pageviews \\
GT resolution 
& Individual-level (daily) 
& Aggregate (daily)
& Aggregate (daily) \\
Supervision method & \makecell[c]{Direct: MOH dates $\to$ \\ SEIRD states $\to$ rates} & \makecell[c]{Hybrid: S\&P 500 indices \\ $\to$ inferred sentiment} & \makecell[c]{Measurement model: \\ Views $\to$ states $\to$ transitions} \\

\midrule
\textbf{Temporal Coverage} \\
Date range 
& Jan 23 -- Apr 14, 2020 
& Jul 1 -- Dec 31, 2024 
& Dec 1, 2024 -- Feb 28, 2025 \\
Total days ($T$) 
& 83 
& 184 (2 financial quarters) 
& 90 \\

\midrule
\textbf{Train-Test Windows} \\
Lookback ($L$) & 28 days & 28 days & 28 days \\
Horizon ($H$) & 7 days & 7 days & 7 days \\
Causality & Strictly causal & Strictly causal & Strictly causal \\
Prediction target & 5 transition rates: & 3 sentiment transitions & 3 attention transitions: \\
& $[S \to E \to I \to R/D \to S]$ & $[Be \leftrightarrow Bu \leftrightarrow N]$ & $[U \leftrightarrow I \leftrightarrow F]$ \\

\midrule
\textbf{Input / Output Dimensions} \\
Tabular features & 55 & 62 & 48 \\
Temporal features & $[28, 12]$ & $[28, 15]$ & $[28, 10]$ \\
Graph embedding dim & 128 & 128 & 128 \\
Output shape 
& $[7, 5]$ 
& $[7, 3]$ 
& $[7, 3]$ \\
Total data samples & 8,352 & 14,500 & 12,890 \\

\bottomrule
\end{tabular}
\end{table*}

\vspace{0.5em}
\paragraph{Modeling Objective.}
To avoid ambiguity, we clarify the scope of the task. The objective is \emph{not} to predict individual agent actions, asset prices, engagement counts, or to optimize pointwise forecasting accuracy on observed signals. Rather, the goal is to recover coherent population regimes, track their temporal evolution, and anticipate regime transitions with calibrated uncertainty.

\vspace{0.25em}
\textbf{Domain-specific objectives:} While the abstract task is shared, the semantic interpretation of latent states and the inference objective differ by domain:

\textit{\textbf{Epidemiology:}} The goal is to infer and forecast the temporal evolution of disease progression at the population level, including the timing and magnitude of transitions between epidemiological compartments. The emphasis is on recovering coherent epidemic dynamics rather than predicting individual infection outcomes in isolation.

\textit{\textbf{Finance:}} The goal is to recover latent market sentiment regimes and their transitions over time, capturing collective belief shifts and contagion effects among traders. The task is explicitly not to forecast asset prices or short-term returns, but to infer the underlying regime dynamics that govern market behavior under uncertainty and high volatility.

\textit{\textbf{Social Attention:}} The goal is to infer collective attention dynamics; emergence, saturation, and decay of engagement states driven by information diffusion. The focus is on modeling attention regimes and their transitions, rather than predicting raw engagement quantities such as pageviews or social media activity.

\vspace{0.5em}
\paragraph{Ground Truth Definition:} The notion of ``ground truth'' varies across domains depending on the observability of latent states.  
In epidemiology, partial ground truth is available through clinical records, enabling deterministic or semi-deterministic reconstruction of latent states from confirmed case timelines.  
In contrast, latent states in the financial and social domains are inherently unobservable at the individual level. Ground truth in these settings is therefore defined via \emph{structured inference procedures} that act as macro-level grounding signals, including rule-based mappings and probabilistic measurement models applied to observed data.  
As a result, evaluation measures alignment with inferred latent regime dynamics rather than fidelity to raw observations.

\vspace{0.5em}
\paragraph{Unified Task Abstraction:} Under this formulation, all datasets instantiate the same abstract inference problem: $\textbf{Observed Signals} \;+\; \textbf{Interaction Structure}
\;\;\longrightarrow\;\;
\textbf{Latent Cluster-Level State Dynamics}$. This unified abstraction enables a single modeling framework to be evaluated consistently across epidemiological, financial, and social systems despite substantial differences in data sources and domain semantics.

\subsection{Dataset Overview}
\label{supp-sec:dataset-overview}

This section provides a high-level overview of the datasets used across domains, summarizing their scale, temporal coverage, latent state structure, and supervision characteristics. Detailed construction procedures for populations, networks, and ground truth inference are deferred to subsequent sections.

\vspace{0.5em}
\paragraph{Summary of Datasets:} All datasets are constructed to support population-level inference of latent regime dynamics under partial observability. Each dataset consists of a heterogeneous agent population, an explicit interaction network, and a sequence of daily observations from which latent states must be inferred at the cluster level.

\begin{table*}[t]
\caption{Summary of experimental domains, agent populations, and temporal scales.}
\label{tab:domain-summary}
\vskip 0.15in
\begin{center}
\begin{small}
\begin{tabularx}{\textwidth}{l @{\extracolsep{\fill}} cccc}
\toprule
\textbf{Domain} & \textbf{Agents ($N$)} & \textbf{Time Span ($T$)} & \textbf{Latent States} & \textbf{Observability} \\
\midrule
Epidemiology (Singapore)    & 1,000 & 83 days  & S, E, I, R, D                & Partial \\
Finance (Market Sentiment)  & 100   & 184 days & Bull, Bear, Neutral          & Latent \\
Social (Climate Attention)  & 250   & 90 days  & Unaware, Interested, Fatigued & Latent \\
\bottomrule
\end{tabularx}
\end{small}
\end{center}
\vskip -0.1in
\end{table*}

\vspace{0.5em}
\paragraph{Temporal Resolution and Regime Dynamics.}
All datasets operate at a \textbf{daily temporal resolution} and are designed to capture non-stationary regime dynamics. Each evaluation period includes both gradual transitions and abrupt regime shifts, such as policy interventions, market shocks, or viral attention events, enabling assessment of temporal alignment, transition detection, and uncertainty calibration.

\vspace{0.5em}
\paragraph{Observed Signals and Supervision.}
Across domains, supervision is defined at the \emph{cluster–time} level rather than the individual agent level. Observed signals consist of aggregate behavioral measurements, contextual exogenous variables, and interaction networks, as described in the Task Overview. With the exception of partially observed epidemiological outcomes, latent states are not directly observable and must be inferred via structured procedures described in Section~X.3.

\vspace{0.5em}
\paragraph{Unified Experimental Framing.}
Despite differences in domain semantics and data sources, all datasets conform to the same experimental framing:
\begin{itemize}
    \item a fixed agent population with heterogeneous attributes,
    \item an explicit interaction graph encoding relational structure,
    \item discrete latent states evolving over time,
    \item and evaluation focused on population-level state dynamics.
\end{itemize}
This unified framing enables consistent evaluation of \textbf{PhysicsAgentABM} across epidemiological, financial, and social systems.

\subsection{Population \& Network Synthesis}
\label{supp-sec:contact-network-synthesis}

This section describes how agent populations and interaction networks are constructed across domains. The procedure is shared across datasets, with domain-specific instantiations deferred to Section~X.4.

\paragraph{Agent Population:} Each dataset consists of a fixed population of $N$ agents, where agents represent individuals, market participants (traders), or information consumers depending on the domain. Agents are heterogeneous, characterized by static attributes (e.g., demographic or behavioral priors) and dynamic latent states that evolve over time.

For \textbf{Epidemiology,} we enrich our 1000 agent profiles using GPT-4o by expanding upon the available demographics, employment, and behavioral attributes of the 1000 patients from the Singapore MOH database. Since the objectives of the simulation for Finance and Social Sciences are to model the population-level emerging patterns, we construct our diverse agent profiles fully synthetically by mimicking complex real-world participants in these settings.  

\paragraph{Interaction Graph:}
Agent interactions are encoded as a weighted graph $G = (V, E)$, where nodes correspond to agents and edges capture domain-relevant relational structure. Edge weights reflect interaction intensity, influence strength, or correlation magnitude, depending on the domain. The interaction graph remains fixed over the evaluation window unless explicitly stated otherwise in the domain-specific sections.

Across domains, interaction networks are constructed to satisfy three shared principles:
\begin{itemize}
    \item \textbf{Structural heterogeneity:} Degree distributions and local connectivity patterns are non-uniform, reflecting realistic interaction asymmetries.
    \item \textbf{Contextual relevance:} Edge semantics align with the dominant interaction mechanism of the domain (e.g., contact, influence, co-movement).
    \item \textbf{Population-scale coherence:} Graph structure supports meaningful aggregation of agent behavior at the cluster level.
\end{itemize}

In all datasets, population synthesis and network construction provide the structural substrate over which latent state dynamics unfold. Differences across domains arise solely from the interpretation of nodes, edges, and weights, and are specified explicitly in Section~X.4.

\subsection{Latent State and Ground Truth Construction}
\label{supp-sec:ground-truth}

This section describes how latent states and supervisory signals are defined across domains. While the semantic meaning of states is domain-specific, the construction procedure follows a shared template.

\vspace{0.4em}
\paragraph{Latent State Space.}
Each dataset defines a small, discrete set of latent states representing population-level behavioral regimes. Latent states evolve over time through stochastic transitions and are inferred at the cluster level rather than assigned to individual agents.

\vspace{0.4em}
\paragraph{Ground Truth Signals:} Ground truth is defined at the population level and varies by domain according to observability constraints.  In epidemiology, partial ground truth is derived from clinical case records, enabling deterministic or semi-deterministic reconstruction of latent states. Specifically, we infer the S,E,I,R,D states using the dates of infection, recovery or death for each patient. An agent is marked \textit{Exposed} 5-7 days before their infection date to account for the disease incubation period, while for dates before this incubation window, they are marked \textit{Susceptible}. This provides us with reliable individual-level latent-state ground truth for Epidemiology. We use the cumulative count of \textit{active infected/recovered/dead cases} as our population-level ground truth. In finance and social domains, latent states are not directly observable; We therefor consider indirect macro signals such as daily S\&P 500 index values and daily Wikipedia page view counts as our ground truth. For \textbf{Finance}, we match every agent's daily latent state with the realized S\&P 500 regime state (Bearish, Bullish or Neutral) to obtain our inference metrics. For \textbf{Attention Diffusion}, we compute normalized scores of daily Wiki page views and compare our normalized attention scores for the agent latent states for quantitative metrics, while for plots, we use attention scores aggregated over our 7 day test windows.

\end{document}